\documentclass[journal,twoside,print]{ieeecolor}
\usepackage{xcolor} 
\RequirePackage{xcolor}
\usepackage{comment} 
\usepackage{generic}
\usepackage{cite}
\usepackage{amsmath,amssymb,amsfonts}
\usepackage{graphicx}
\usepackage{textcomp}
\usepackage{hyperref}
\usepackage{booktabs}
\usepackage{multirow}
\usepackage{algorithm}
\usepackage{algpseudocode}
\usepackage{comment}
\usepackage{svg}
\usepackage{tabularx}
\usepackage[table]{colortbl}
\usepackage{tikz}
\usepackage{tikzpeople}
\usepackage{subcaption}
\usepackage{etoolbox}
\usepackage[aboveskip=0pt,belowskip=-1pt]{caption}
\makeatletter
\@ifundefined{color@begingroup}%
  {\let\color@begingroup\relax
   \let\color@endgroup\relax}{}%
\def\fix@ieeecolor@hbox#1{%
  \hbox{\color@begingroup#1\color@endgroup}}
\patchcmd\@makecaption{\hbox}{\fix@ieeecolor@hbox}{}
\patchcmd\@makecaption{\hbox}{\fix@ieeecolor@hbox}{}

\usetikzlibrary{fit}

\def\BibTeX{{\rm B\kern-.05em{\sc i\kern-.025em b}\kern-.08em
    T\kern-.1667em\lower.7ex\hbox{E}\kern-.125emX}}
\begin{document}
\title{Transfer Learning with Active Sampling for Rapid Training and Calibration in BCI-P300 Across Health States and Multi-centre Data}
\author{Christian Flores, \IEEEmembership{Member, IEEE}, Marcelo Contreras, Ichiro Macedo and Javier Andreu-Perez, \IEEEmembership{Senior Member, IEEE}
\thanks{This work was partially supported by Fondo Semilla, UTEC, under Grands 871059-2022.}
\thanks{Christian Flores is with the Centro BIO UTEC and Department of Electrical and Mechatronic Engineering, UTEC, Lima, Peru. Also, He is with DSPCom Laboratory from the Department of Electrical Engineering, University of Campinas, Campinas, Brazil (e-mail: cflores@utec.edu.pe).}	
\thanks{Marcelo Contreras and Ichiro Macedo are with the Department of Electrical and Mechatronic Engineering, UTEC, Lima, Peru.}
\thanks{ Javier Andreu-Perez is with the Centre for Computational Intelligence, University of Essex, Colchester, United Kingdom} }

\maketitle
\begin{abstract}
% Referencia:
%  https://arxiv.org/pdf/2107.04470.pdf
%  https://personal.ntu.edu.sg/ctguan/Publications/2022_Luzheng_IEEE_TBME(EA).pdf_
%
Machine learning and deep learning advancements have boosted Brain-Computer Interface (BCI) performance, but their wide-scale applicability is limited due to factors like individual health, hardware variations, and cultural differences affecting neural data. Studies often focus on uniform single-site experiments in uniform settings, leading to high performance that may not translate well to real-world diversity. Deep learning models aim to enhance BCI classification accuracy, and transfer learning has been suggested to adapt models to individual neural patterns using a base model trained on others' data. This approach promises better generalizability and reduced overfitting, yet challenges remain in handling diverse and imbalanced datasets from different equipment, subjects, multiple centres in different countries, and both healthy and patient populations for effective model transfer and tuning.

In a setting characterized by maximal heterogeneity, we proposed  P300 wave detection in BCIs employing a convolutional neural network fitted with adaptive transfer learning based on  Poison Sampling Disk (PDS)  called Active Sampling (AS), which flexibly adjusts the transition from source data to the target domain. Our results reported for subject adaptive with \(40\%\) of adaptive fine-tuning that the averaged classification accuracy improved by \(5.36\%\) and standard deviation reduced by  \(12.22\%\)  using two distinct, internationally replicated datasets. These results outperformed in classification accuracy, computational time, and training efficiency, mainly due to the proposed Active Sampling (AS) method for transfer learning.

 % 83.49-78.13
% 17.44-5.22 STD

%Our model achieved an average classification accuracy of $81,07 \pm 8,30$ (Avg. $\pm$ SD) for subject adaptive with \(40\%\) of adaptive fine-tuning using two distinct, internationally replicated datasets and outperformed in both accuracy, speed, and training efficiency, largely due to the proposed method PDS for transfer learning.

\end{abstract}


\section{Introduction}
\label{sec:introduction}

\IEEEPARstart {A} Brain-Computer Interface (BCI) decodes brain signals for communication between a human and their environment \cite{Villa2021}. EEG signals are preferred for being non-invasive and capable of recording specific brain activities like motor or mental tasks, thereby extracting important information \cite{Villa2021}. They offer a balanced spatio-temporal resolution through a high-density electrode system, improving BCI performance \cite{Villa2021}. The P300 BCI paradigm produces a P300 wave in EEG signals via the oddball method. This wave, an event-related potential (ERP), reacts to visual or auditory stimuli \cite{Villa2021}. %For instance, displaying a sequence of images and flashing between the image transitions can trigger a peak in ERP called P300, which manifests as a positive voltage peak after 300 ms of stimulus exposition \cite{Farwell1988}.
%----- Inicio: Borrado por la reducción de páginas-------%
%BCI's classification stage had numerous Machine Learning (ML) proposals, producing suitable classification accuracies. In \cite{HOFFMANN2008115}, Bayesian Linear Discriminant Analysis (BLDA) achieved 100\% accuracy by increasing the number of blocks for healthy and disabled subjects. Single-channel EEG recordings of emotions were classified with Support Vector Machines (SVM) and $k$-Nearest Neighbors algorithm (k-NN) in \cite{Jalilifard}, reporting  92\% and 94\% accuracy, respectively. In \cite{Srinivasan}, Elman Network (EN) and Probabilistic Neural Network (PNN) classified seizure EEG signals with Approximate Entropy (AEn) as the only feature, achieving an overall 99.3\% $\sim$ 100\% accuracy by EN and 98\% $\sim$ 100\% by PNN. Feature extraction is an important stage in ML classification to enhance the accuracy classification score, yet it can be tedious and sensitive. %, a crucial drawback of ML solutions.
%--------------------
%----- Fin: Borrado por la reducción de páginas-------%

Deep Learning (DL) has been effective in the classification stage of BCIs, especially through Convolutional Neural Networks (CNNs) and their variants, due to their proficiency in extracting spatial and temporal features from EEG signals. For example, Yang et al. \cite{Yang2020} achieved 86.41\% accuracy in classifying the BCI Competition IV dataset using a multi-layer CNN. Vega et al. \cite{Vega2022} developed EEG-TCFNet, combining EEGNet with temporal convolution layers and fuzzy blocks, reaching up to 98.6\% accuracy for subject-dependent P300 wave classification. However, DL models require large, homogeneously distributed datasets for reliable accuracy and generalization, which is challenging given the laborious nature of EEG experiments. To address this, some studies have implemented under-sampling methods to balance datasets, enhancing classification accuracy \cite{Cortez_2020_SSCI, Cortez_2020_SMC}.

%Additionally, subject-dependent independent strategies which use data of all subjects for training to tackle data shortage may not give consistent results for all subjects. 
% Despite, few examples of DL research were mentioned above, they all have one thing in common: they focused on classifying a particular BCI job while employing detailed information of that task to create the network architecture. Also, it should be emphasized that the size of the database used to train the various networks differed depending on the study, making it difficult to generalize these deep learning techniques to other BCI tasks or databases with different size. Even, the number of samples between classes can vary, so datasets have an unbalanced class distribution that could penalize deep learning models.
% It is worth to mention that the problem of data collection for new individuals as well as the issue of accuracy in the classification of EEG signals that exists in the field of BCIs as a whole have not been resolved in the aforementioned researches. When collecting data from fresh BCI users, it is preferred to use the fewest quantity of EEG signals that can be obtained.
Transfer Learning (TL) schemes in DL models address data variations from different subjects, sessions, and devices, enhancing performance. TL involves pre-training DL networks on large datasets to create generalized models with high performance on new data \cite{liu_2018, Bird2020, Xu2019, 8857575}. For example, models like Residual Neural Network (ResNet) and VGG16 pre-trained on the ImageNet dataset have been used for mental task signal classification, reaching accuracies up to 86.85% \cite{Singh2020}.

The fine-tuning aspect of Transfer Learning from source to target domains presents challenges, including the risk of negative transfer, which is difficult to manage. This process necessitates a judicious selection to prevent irrelevant source data from affecting the target domain mapping. \emph{Adaptive Transfer Learning (ATL)}, as proposed in \cite{Zhang_Guan_2021}, addresses this by adaptively sampling the target domain for mapping. While ATL mitigates inter-subject variability and manages unbalanced datasets, it incurs high computational costs and struggles with intra-subject variability.

Optimizing Stochastic Gradient Descent (SGD) parameters is key to enhancing classification accuracy and convergence speed, primarily by reducing Stochastic Gradient Noise (SGN) for quicker convergence and larger learning rates. As noted in previous studies, Point Processes (PP) can effectively reduce SGN by creating diverse, non-redundant mini-batches for DL model training \cite{Peilin_2014}. However, this method might repeatedly select the same samples. To counter this, we introduced Active Sampling (AS), which samples the entire dataset to form a new dataset for DL model training. Active Sampling improves dataset efficiency and diversity, enhancing SGD accuracy and convergence, and when combined with ATL, it effectively addresses data variability for subject-independent classification. %!!!!We tested our framework in two scenarios: original and extended multi-center experiments for dependent, independent, and adaptive classification.

The \textbf{highlights} of this work are:

\begin{enumerate}
    \item \textit{\textbf{We propose Active Sampling (AS), a new method based on Poison Disk Sampling (PDS) for Adaptive Transfer Learning (ATL)}} to improve adaptive fine-tuning in deep learning models for P300 BCI decoding tasks. AS efficiently samples source and target subjects to expedite training and calibration in the ATL framework.
    %\item A new improved method for adaptive fine-tuning in deep learning models in a P300 BCI decoding task based on the Repulsive Point Process (RPP) of the target domain data.
    %\item An analysis is performed in two replicas of a BCI experiment carried out in two completely different locations (multi-centre).% and different materials.
  
    \item  \textit{\textbf{To evaluate the proposed approach, we devised a new extended experimental scenario}}. We replicated an experimental dataset used as a benchmark for P300 but with different equipment, subjects, and centres to introduce real-world variability. Both datasets were combined to test the proposed AS-based method's resilience. 
    %\item For experimentation, data from health as well as patients (stroke) were used for evaluation.
    
    %\item An assessment of Adaptive Transfer  Learning  (ATL) is performed in subject specific and cross-subject specific settings.
    
   %\item A notable gain has been achieved for signal classification on the motor imagery BCI dataset in various scenarios compared to other end-to-end architectures.
    \item The implementation of AS within the P300 BCI paradigm has resulted in a \textit{\textbf{substantial reduction in performance variability and computational expense. This approach underwent rigorous testing in various classification contexts}}, including subject-dependent, subject-independent, and subject-adaptive scenarios, yielding 
\end{enumerate}

%The assessment of inclusion of mini-batch sampling by Repulse Point Processes in P300 EEG BCI achieve remarkable enhancement on accuracy score and convergence speed, cross-subject and subject specific. 
% Puede

%achieves remarkable enhancement in 
%for cross-subject and subject-specific.

\begin{comment}
\begin{itemize}
    \item An assessment of inclusion of mini-batch sampling by Repulse Point Processes in EEG signal classification to achieve remarkable enhancement on accuracy score and convergence speed. The PSD algorithm chooses \textcolor{red}{highly correlated point samples} and the ones with the most relevant information on feature space between classes. Correlation between samples was measured using Frobenius distance.     
    \item An adjustment of the adaptive TL schemes proposed in \cite{Zhang2021} to classify P300 stimulus instead of MI signals.     
\end{itemize}
\end{comment}

The rest of the paper is organized as follows: Section \ref{related} presents related previous works, Section \ref{background} describes the state-of-the-art of sampling methods,  Section \ref{methodology} introduces methodology including dataset details and classifying pipeline, Section \ref{results} reports the results and Section \ref{discussion} discusses our results and their importance. We finally conclude the paper in Section \ref{conclusion}.

% Se podria aumentar 100 palabras
\section{Related work} \label{related}

This section reviews studies relevant to our proposal. Pan et al. \cite{Pan_2021} proposed transfer learning to address the issue of learning with insufficient labeled data, primarily by leveraging knowledge from related domains. A sampling strategy based on distribution difference is developed to select the most valuable instances for label querying. Liu et al. \cite{Liu_2023} employed transfer learning and active sampling methods to tackle imbalanced data in classification. The proposed transfer learning model comprises three modules such: an active sampling module, a real-time data augmentation module, and a DenseNet module.
%This section reviews studies relevant to our model, 
Some studies focus on how Stochastic Gradient Noise (SGN) acts as a regularizer in deep learning and its role in improving generalization through Stochastic Gradient Descent (SGD) \cite{Xie_2021}. Key studies by Kulezka and Taskar \cite{Kulesza_2012_a, Kulesza_2011, Kulesza_2012_b} explore \emph{Determinantal Point Processes (DPP)} in subset selection for data diversification in machine learning applications. Zhdanov \cite{Zhdanov2019} proposed an algorithm for creating diverse mini-batches, emphasizing informativeness and sample spacing, and incorporating \textit{k}-means for scalability. Additionally, Point Processes (PP), like DPP or \emph{Repulsive Point Processes (RPP)}, are used for uncorrelated feature sampling in mini-batches, enhancing accuracy and convergence speed \cite{Zhang2017, Zhang2018}.
%#----- Fin
Zhang et al. \cite{Zhang2017} utilized DPP to lower the variance in SGD. They applied a non-uniform DPP-based sampling method to eliminate redundant data in mini-batches, demonstrating faster DPP-SGD convergence compared to the standard SGD on CIFAR-10 and Oxford-102 datasets. Similarly, Zhang et al. \cite{Zhang2018} proposed using an RPP, specifically Poison Disk Sampling (PDS), for mini-batch sampling to simplify the RPP's complexity. Their studies compared SGD baseline, DPP-SGD, and PDS-SGD, revealing a significant reduction in convergence time with PDS across various tests.
As discussed, although several studies have approaches to classify the target action based on the P300 BCI, some issues need to improve:
\begin{itemize}
\item  Reduce the calibration time of BCI operation to operate in real-life conditions.
\item  Improve the model decoding capabilities to enhance classification accuracy, reducing the computational costs during the calibration stage.
\end{itemize}

%\textcolor{blue}{These issues are notable hurdles to face, especially in scenarios where the diversity of electrophysiological signals across different individuals, experiments conducted at varied locations, and utilizing diverse materials. Thus, Active Sampling (AS) samples relevant information from different distributions, eliminating redundant or spurious samples from the EEG datasets. It paves the way for better Adaptive Transfer Learning (ATL) performance to expedite the calibration time of BCI operation for new subjects, which is appropriate for implementing BCIs in real-world scenarios.} 

These issues are notable hurdles to face, especially in scenarios where the diversity of electrophysiological signals across different individuals, experiments conducted at varied locations, and utilizing diverse materials. Thus, Active Sampling (AS) paves the way for better Adaptive Transfer Learning (ATL) performance to expedite the calibration time of BCI operation for new subjects, which is appropriate for implementing BCIs in real-world scenarios. The proposed method resulted in enhancements for both subject-dependent and subject-independent settings, covering both healthy individuals and patients, and was applied across two experimental datasets.

%% Se podria aumentar 50 palabras 

\section{Background} \label{background}%; Debe fusionar con intro o related work}
% 470 palabras
This section summarizes the state-of-the-art of RPP and PDS as the chosen sampling method. In addition, we also delineated that RPP may reduce the variance of SGD. Therefore, in the present work, we included RPP to sample the data from P300 signals to reduce the calibration Process in P300 BCI.

\subsection{ Repulsive Point Processes (RPP) }
In this section, we provide a formal definition of the  RPP. Considering  a point process $P$ in $R^d$ and the $n^{th}$ order product density $\varrho^{(n)}$ is defined by:
\begin{equation}
p(x_1,....,x_n)=\varrho^{(n)}(x_1,x_2,....,x_n)dx_1....dx_n
\label{Main_eq}
\end{equation}
where $p(x_1,....,x_n)$ is the joint probability of having a point of the Point Process P in each of the infinitesimal spheres B. For our analysis, we define  the first \ref{first_eq}   and second \ref{second_eq} order product density, which are commonly denoted by:
\begin{equation}
\lambda(x) := \varrho^{(1)}(x)
\label{first_eq}
\end{equation}
\begin{equation}
\varrho(x,y) := \varrho^{(2)}(x,y)
\label{second_eq}
\end{equation}
Furthermore, to denote the gradient of the loss function, it can be expressed as $g(x,\theta) = \nabla(x,\theta)$  and $k=\lvert B \rvert $, the mini-batch size. After mentioning the previous definition, we can define:

\textbf{Theorem 1}. The variance of the gradient $var_{P}(\hat {\boldsymbol{G}})$ estimate $(\hat {\boldsymbol{G}})$ in SGD for a general stochastic point process P is given by:
\begin{equation}
\begin{aligned}
var_{P}(\hat {\boldsymbol{G}})=\frac{1}{k^2} \int_{\nu \times \nu } \lambda(x)\lambda(y) \boldsymbol g(x,\theta)^{T}\boldsymbol g(y,\theta) \\
[\frac{\varrho(x,y)}{\lambda(x)\lambda(y)}-1]dxdy  \\
+\frac{1}{k^2} \int_{\nu} \left\|{g(x,\theta)}\right\|^{2}\lambda(x)dx
\end{aligned}
\label{SGD_equation}
\end{equation}
Thereby, the RPP can make the first term in Eq. \ref{SGD_equation} negative and reduce the variance.
\\
\begin{proof}
For repulsive point processes, the probability of sampling points that are close to each other is low.
Hence, if $x$ and $y$ are close, then correlations $\varrho(x,y)<\lambda(x)\lambda(y)$ make the cocient \( \left[\frac{\varrho(x,y)}{\lambda(x)\lambda(y)}\right] \) tend to zero, and as $-1$, negative. Furthermore, supposing that the loss function is sufficiently smooth in its data argument, the gradients are aligned for close points, i.e., $\boldsymbol g(x,\theta)^{T}$, $\boldsymbol g(y,\theta)$. Thus, close points provide a negative contribution to the first integral in Eq. \ref{SGD_equation}, so the negative first term in this equation drives the process to variance reduction.
\end{proof}

%\subsubsection{Poison disk sampling (PSD)}
\subsection{ Poison Disk Sampling (PDS) }

PDS is one type of RPP that reported stronger local repulsion than DPP. A dart-throwing algorithm is used to implement a PDS and provides equivalent point arrangements to DPP but much more efficiently. \cite{Lagae_2008}. This algorithm states that the smallest distance between each pair of sample points should be at least to the predefined distance $(r)$. Thus, when the distance between two points is smaller than the disk radius 
$\left\|{x-y}\right\|\leq r$, the second order product density $\varrho(x,y)$ for PDS is zero, and when the two points are far, the second order product density converges to $\varrho(x,y)=\lambda(x)\lambda(y)$. Moreover, the complexity of PDS is $\mathcal{O}(k^2)$, much lower than DPP complexity $\mathcal{O}(Nk^3)$ where N is the total amount of points. Additionally, its simple procedure allowed the proposal of numerous PDS variants and spread the sampling even more.

\section{Methodology} \label{methodology}
In this section, we detail EEG signal preprocessing and data sampling using PDS. We then describe the process of training a Deep4Net network for three scenarios: subject-dependent (SD), subject-independent (SI), and subject-adaptive (SA), applying ATL to the pre-trained model. This methodology was replicated in two different countries using distinct equipment but maintaining the same paradigm.
%The following scheme \ref{fig:flowchartwear} illustrates in more detail the presented pipeline, and this section is going to detail its key components. 

\subsection{Datasets}
%\label{methodology}
\label{sec:methodology}
%We evaluated our scheme proposal in two scenarios. 
We employed two datasets: the public `Original Experimental benchmark dataset clinical P300 dataset' (OE) and our own `Multi-centre benchmark clinical P300 dataset' (ME), created in our lab. ME replicates OE's P300 recording protocol but differs in device use, subjects, patients, and locations. Detailed descriptions of each dataset will follow in subsequent subsections.
%that merges the dataset from the original experiment plus the replication study of the same P300 conditions but uses different device acquisition, subjects, patients, locations, etc. 
\subsubsection{Original experimental benchmark clinical P300 dataset (OE)} 
This dataset comprises P300 recordings from nine individuals, including four disabled patients (S1 to S4) and four healthy PhD students (S5 to S9), though data from S5 was excluded due to communication issues during the experiment. Demographic details are in Table \ref{Hoffman_dataset_table}. EEG signals were captured using a Biosemi Active Two amplifier with 32 channels at a 2048 Hz sampling rate, following the 10-20 system. The EEG recording involved showing six images in random sequences on a screen, with each image flashing for 100 ms followed by a 300 ms pause. Each participant completed four sessions, each containing six runs with 20-25 blocks chosen randomly. Further procedural details are available in \cite{HOFFMANN2008115}. %This experimental procedure was the same for UTEC dataset.
\begin{table}[!h]
\scriptsize
\caption{Demographic information of subjects of OE}
    \centering
    \begin{tabular}[b]{l l l l}
    \toprule
\textbf{Sub.} & \textbf{Age} & \textbf{Gender} & \textbf{Diagnosis}  \\
        \midrule
      S01  & 56 & Male & Cerebral patsy \\
      S02  & 51 & Male & Multiple sclerosis \\
      S03  & 47 & Male & Late-stage amyotrophic \\
      S04  & 33 & Female & Traumatic brain and spinal-cord \\
      S05  & 43 & Male & Post-anoxic encephalopathy\\
      S06  & Around 30 & Male & Healthy \\
      S07  & Around 30 & Male & Healthy \\
      S08  & Around 30 & Male & Healthy  \\
      S09  & Around 30 & Male  & Healthy  \\
\bottomrule
\end{tabular}
        
    \label{Hoffman_dataset_table}
\end{table}

%\hl{Revisar numeración de sujetos considerando que el sujeto 5 fue retirado}

\subsubsection{Multi-centre experimental benchmark clinical P300 dataset (ME)}
This dataset involved nine participants who underwent the Speller test. Table \ref{UTEC_dataset_table} provides additional subject details, with healthy participants being undergraduate students serving as the control group for the patient cohort. Notably, S07 and S08 exhibited mild aphasia, with S08 additionally experiencing upper limb paresis, while S09 had severe apraxia. Ethical approval and participants' written consent were obtained from the Universidad Peruana Cayetano Heredia's ethics committee, ensuring subjects' awareness of the academic objectives, procedures, and anonymity. EEG signals were acquired using a 16-channel g.USBamp amplifier at a 2400 Hz sampling rate, with bipolar electrodes placed according to the 10-20 system. Each subject completed four recording sessions, each comprising six runs with 20-25 randomly selected blocks. For a more detailed procedure description, refer to \cite{UTEC_dataset}.
\begin{table}[!htbp]
\scriptsize
        \caption{Demographic information of subjects of ME}

    \centering
    \begin{tabular}[b]{l l l l}
    \toprule
\textbf{Sub.} & \textbf{Age} & \textbf{Gender} & \textbf{Diagnosis}  \\
        \midrule
      S01  & 33 & Male & Healthy \\
      S02  & 21 & Male & Healthy \\
      S03  & 20 & Male & Healthy  \\
      S04  & 21 & Male & Healthy    \\
      S05  & 24 & Male & Healthy   \\
      S06  & 29 & Male & Healthy  \\
      S07  & 20 & Male & Hemorrhagic post-stroke    \\
      S08  & 52 & Female & Ischemic post-stroke   \\
      S09  & 55 & Male  & Ischemic post-stroke  \\
\bottomrule
\end{tabular}
    \label{UTEC_dataset_table}
\end{table}
%Therefore, in this work two approach was performance for both dataset in the following way: 
%\subsubsection {Original experiment} 
 %we used only a Hoffman dataset to apply different classification scenarios.
%\subsubsection {Extended multi-center experiment}
%In this case, we join a Hoffman dataset and UTEC dataset to apply different classification scenarios in order to enhance accuracy scores using different multi-center samples belonging to each subject
%Hence, this study proposes a generalized Active Sampling (AS) framework to leverage P300 data across multiple domains, such as %sessions, subjects, and acquisition devices to use
% BCI in real-life applications.
%, so we create two analysisscenarios:
%\begin{enumerate}
     %\item Analysis 1 (OE): only use data from the Original Experimental benchmark dataset clinical P300 dataset (OE).
     %\item Analysis 2 (OE+IE): use data from Original Experimental benchmark dataset clinical P300 dataset (OE) combined with the Multi-centre benchmark clinical P300 dataset (IE).
%\end{enumerate}

\begin{figure*}[t!]
    \centering    
    %\includesvg[width=0.75\textwidth]{Figures/Overview_v2.svg}
    \includegraphics[width=0.75\textwidth]{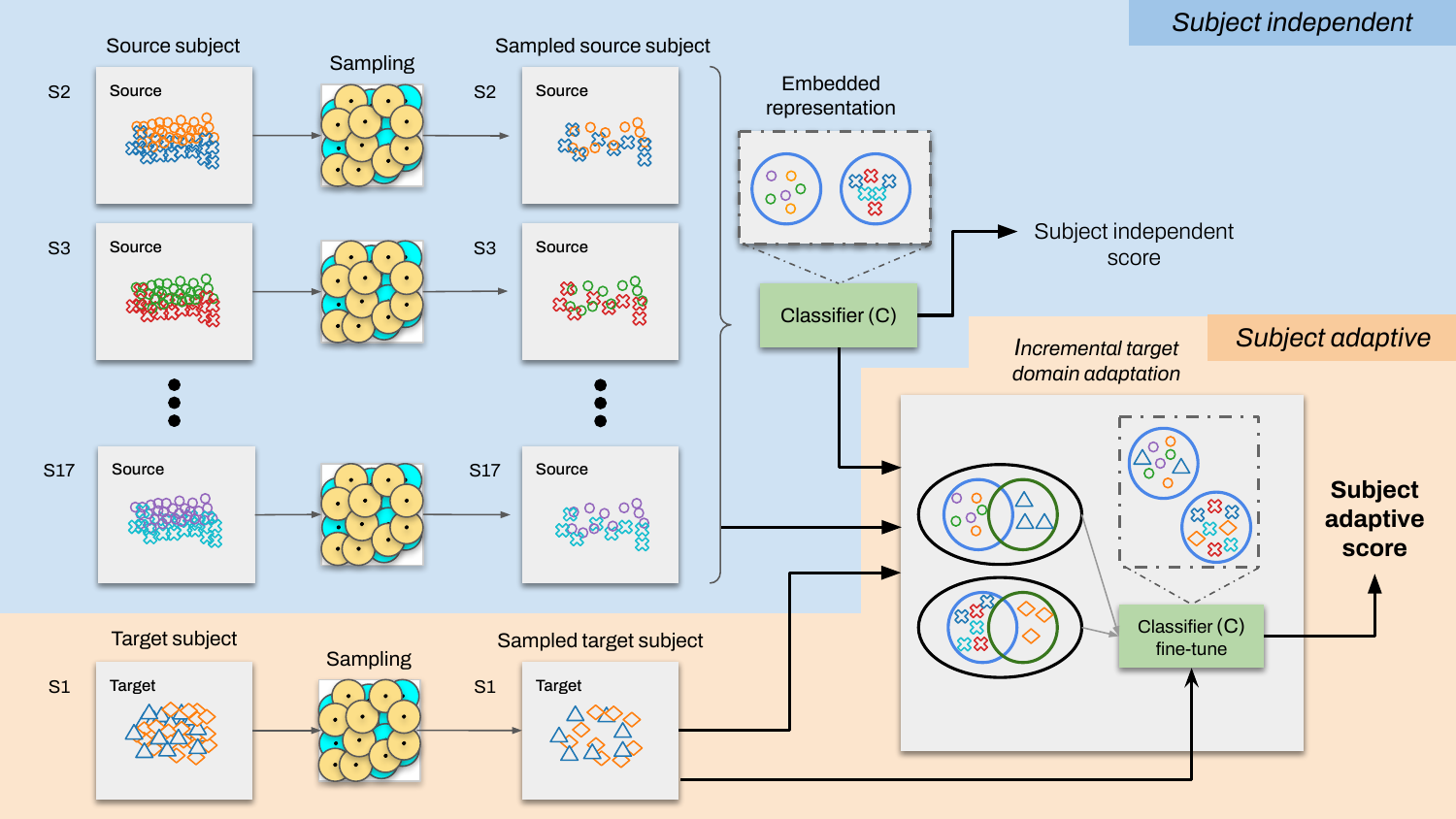}
    \caption{In the present work, we validate the efficacy of adding Active Sampling (AS) by testing the EEG-based P300 decoding task-based using adaptive fine-tuning in deep learning models for subject 1 as the target subject and subjects 2 to 17 as source subjects, which represent the OE+ME w/ AS experimental scheme. The source and target subject are sampled using the AS to diversify and reduce the non-redundant data. The subject-independent and subject-adaptive schemes are represented by different colors in this figure. They and the arrows demonstrate the different blocks considered when computing the adaptive fine-tuning in deep learning models in a P300 BCI for train and testing}
    \label{fig:overview}
\end{figure*}
\subsection{Data pre-processing}\label{Data_pre_processing}
In order to eliminate artifacts and obtain a higher number of samples, the two datasets were preprocessed. As both followed the same steps, only one kind of preprocessing task was designed:

\begin{enumerate}
    \item It applies a sixth-order Butterworth filter between 1 Hz and 15 Hz frequencies.
    \item Invalid spectral components were eliminated using Notch filters.
    \item Signals are downsampled from 2048 Hz to 32 Hz to reduce computational cost.
    \item Elimination of outlier values followed the winsorization criteria.
    \item Signals are windowed with a length of 1000 ms, which overlaps 600 ms because the stimulus interval lasts 400 ms. 
\end{enumerate}

\subsection{Dense Poison Disk Sampling for Active Sampling}
When dealing with highly structured data, there may be only a few data points near the decision boundary that are difficult to classify, such as waves labeled P300 and non-P300. To tackle this issue, we use one PDS variation, namely Dense PDS, which draws darts based on mingling indices instead of random selection. This allows us to specify a categorical distribution called $\pi$ to refine the decision boundary. The mingling index measures the ratio of points belonging to different class labels. For further information on the mingling index, see \cite{Zhang2018}. 
% Borrado por la reducción
%In addition, Fig. \ref{fig:PSD_sampling} illustrates the algorithm and its rejection disk with radius $r_0$  concerning some accepted points. The rejection area ensures sufficient sample spreading. 
Furthermore, a pseudo-code of the procedure is presented in Algorithm 1.

% Borrado por la reducción
%\begin{figure}[!h]
%    \centering
%    \includegraphics[scale = 0.4]{Figures/minibatch.eps}
%    \caption{\textbf{Sampling process through Poisson repulsive disk}: The radius $r_0$ (black circles) defines the rejection disk with respect to a sampled point (red dots). Otherwise, if the new sampled point falls inside of any circles, this point will be rejected}

%   \label{fig:PSD_sampling}
%\end{figure}

In this study, we applied a PDS over the raw data of each subject for each dataset in order to reduce the number of samples using a Sampling Factor (SF). This factor was analyzed experimentally for different $k$ sampling sizes and selected, which yielded better performance. 

\begin{algorithm}[!h]
\footnotesize
\caption{Dense PDS for Active Sampling}\label{alg:kpsd}
\hspace*{\algorithmicindent} \textbf{Input:} $k$ sampling size, $r_0$ rejection radio, $M$ mingling index, $\pi_0$ parameter for categorical distribution to sample mingling index \\
\hspace*{\algorithmicindent} \textbf{Output:} $\mathcal{B}$ active sampling 
\begin{algorithmic}
\State count $\gets$ 0
\While{count $\neq k$}
    \State Sample a mingling index m $\sim$ Cat($\pi$)
    \State Randomly sample a point with mingling index $m$
    \State distance $\gets$ Euclidean norm
    \If{distance $\leq r_0$}
        \State Reject point
    \Else
        \State Append new point in $\mathcal{B}$
        \State count $\gets$ count + 1
    \EndIf
\EndWhile
\end{algorithmic}
\end{algorithm}

\vspace{-0.6cm}

\subsection{Deep Network Architecture}
We utilised Deep4Net CNN architecture as base architecture \cite{Schirrmeister_2017}. The accuracy classification and Cross entropy loss value were performed using AdamW optimizer, Batch Normalization, and  Dropout. The training parameters comprehend a batch size of 16 samples, performing the stage for 200 epochs.

\vspace{-0.3cm}
\subsection{Classification scheme for Active Sampling}
In this section, we perform the effect of Active Sampling (AS) on the two datasets in different classification schemes, such as the subject-dependent, independent, and adaptive classification, using a Deep4Net \cite{Schirrmeister_2017}. %As described in the previous section,
The data for each subject is composed of four sessions; thus, we sampled each subject for both datasets using a Dense PDS described in the previous item.

%475
\begin{figure}[h!]
 \centering
  \includegraphics[width = 0.40\textwidth]{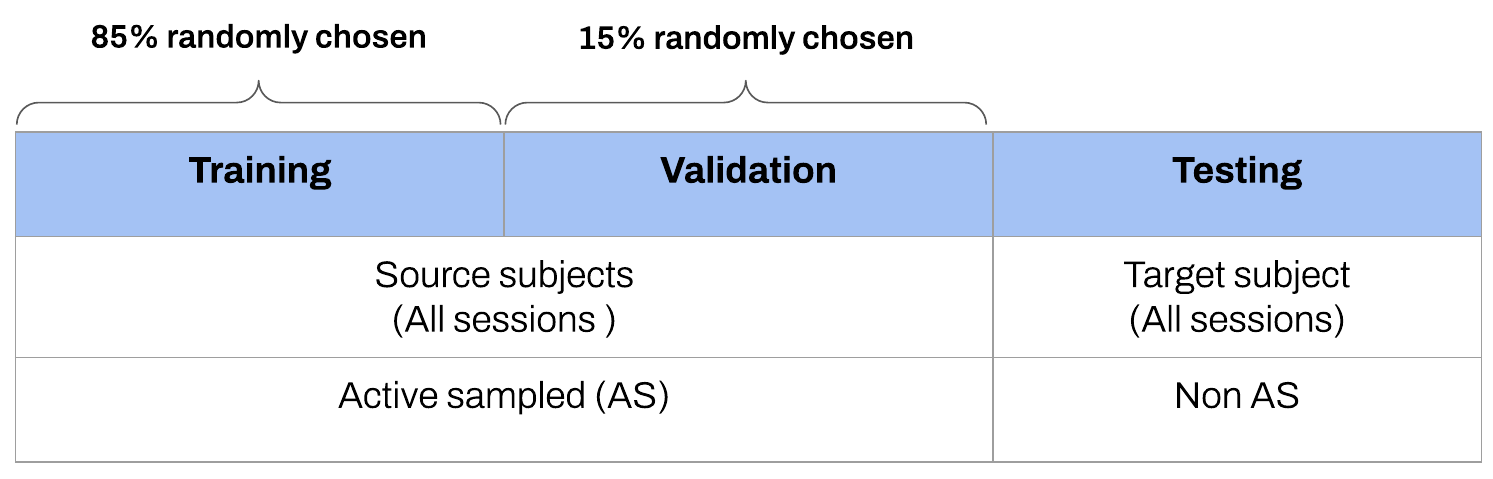}
  \caption{Scheme independent data splitting}  \label{fig:Scheme_independent_portions} 
\end{figure}

\subsubsection {Subject-depended classification}
In this type of classification, we trained and validated a classification model for each subject using only his data. Regarding the Active Sampling (AS), each dataset subject (OE and ME) was sampled for sessions 1, 2, and 3 to the training and validation stage. However, session 4 was not sampled since it was used to test.

\subsubsection {Subject-independent classification}
We used a Leave-One-Subject-Out Cross-Validation (LOSOCV)  to perform this classification scheme so that all data from an independent, also called target subject, is held out as the test set. Concerning the Active Sampling (AS), session 4 of the target subject was kept from being sampled since it was used in the testing stage. However, the rest of the subjects, also called source subjects, either for the OE or the IE dataset, were sampled for all sessions to the training and validation stage. As depicted in the figure \ref{fig:overview}.
\begin{comment}
\begin{figure}[h]
  \centering
  \includesvg[width = 0.5\textwidth]{Figures/1.Esquema Independent.svg}
  %\caption{Multi center expansion results}
    \caption{Scheme Independent}
    \label{fig:Scheme_independent} 
\end{figure}
\end{comment}
In subject-independent classification, we randomly split the data from non-target subjects into 85\% for training and 15\% for validation, as we can see in Fig.\ref{fig:Scheme_independent_portions}

\subsubsection {Subject-adaptive classification}
\begin{figure}[h]
\centering
    \includegraphics[width=0.5\textwidth]{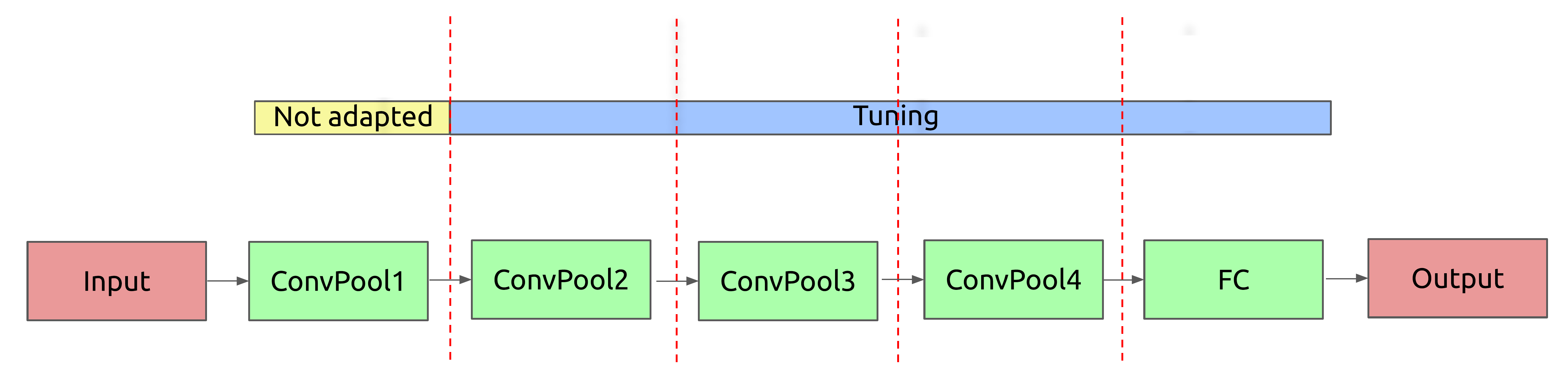}
    \caption{Adaptive learning schemes contrasted with network layers}
    \label{fig:Adaptive_learning_schemes}
\end{figure}

\begin{figure}[h]
\centering
\includegraphics[width = 0.4\textwidth]{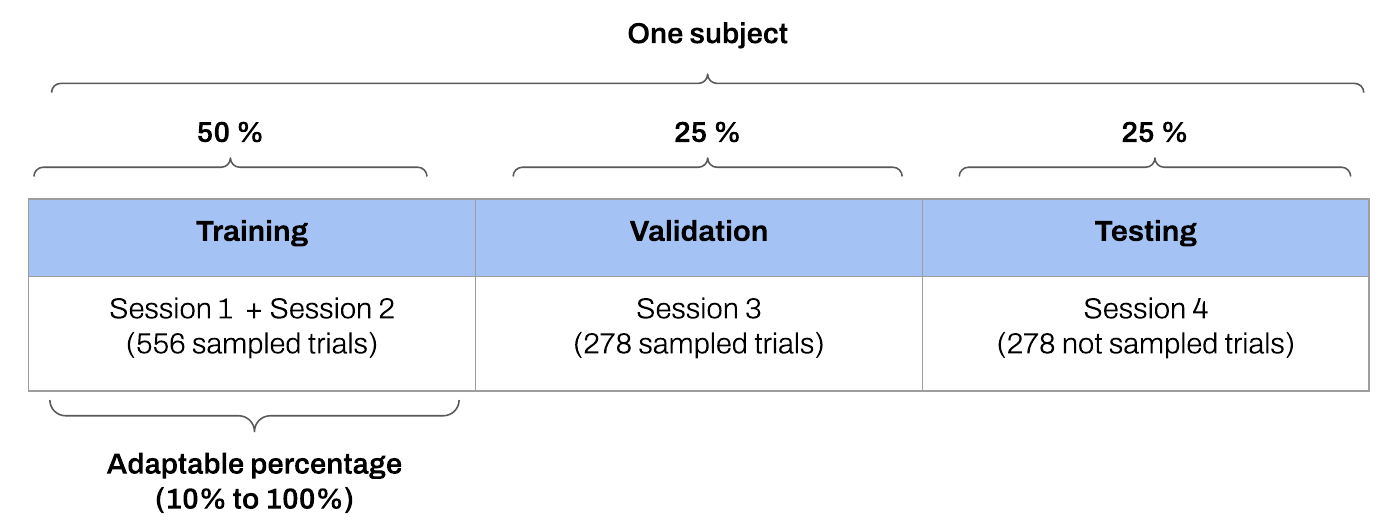}
\caption{Adaptive scheme of the subject target}
\label{fig:Split_data_Adaptive}
\end{figure}

\begin{figure}[h!]
    \centering    
    \includegraphics[width=0.45\textwidth]{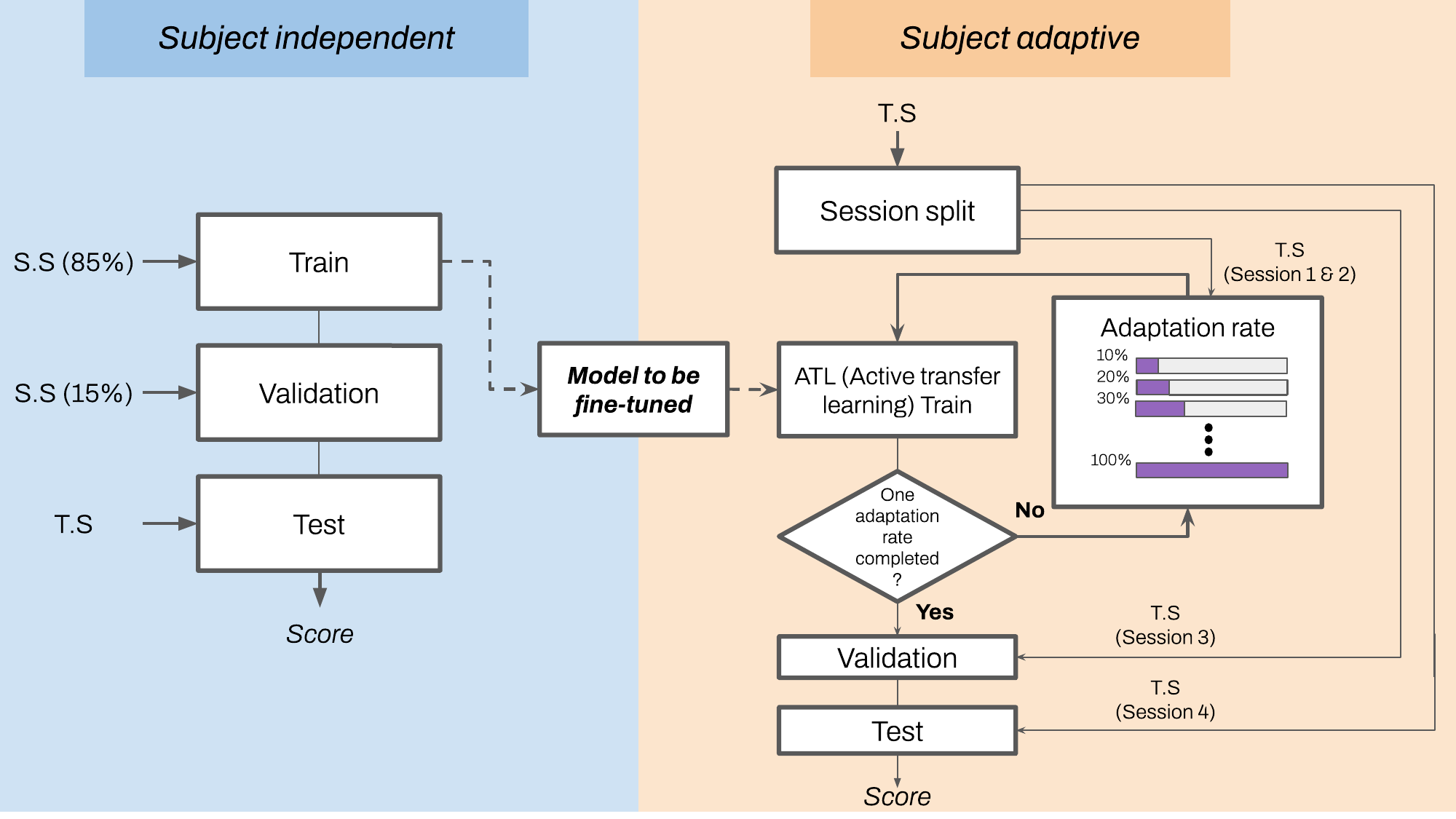}
    \caption{Interconnection between subject independent and adaptive with inputs of \textbf{Source} subjects (S.S) and \textbf{Target} subject (T.S)}
    \label{fig:subject_ind_and_Adap}
\end{figure}

In this study, we implemented a subject-adaptive classification scheme that did not initially include data from the target subject, which can impact accuracy due to variations among subjects. To address this, we fine-tuned a pre-trained model with a small dataset from the target subject, specifically by freezing the first convolutional layer of Deep4Net and adjusting the remaining layers \cite{Zhang_Guan_2021}, as shown in \ref{fig:Adaptive_learning_schemes}.
This model, initially trained in a subject-independent manner, then served as the base for each target subject. We observed performance enhancements as the adaptation rate from the target subject increased in 10\% increments, up to 100\%. Regarding Active Sampling (AS), we utilized the first two sessions for training and the third for validation, while the fourth session was reserved for testing, as depicted in Figure \ref{fig:overview}. 
Moreover, in Fig. \ref{fig:Split_data_Adaptive}, we represent the percentage of data corresponding to the number of sessions for training (50\%), validation (25\%), and testing or evaluation (25\%).
%0.4
Fig. \ref{fig:subject_ind_and_Adap} provides a flowchart with further explanation to better illustrate the difference between subject-adaptive and subject-independent classification.

Therefore, to validate the performance of the Active Sampling (AS)  over the three classification schemes, we used a Deep4Net as described above. Thus, we designed four experimental schemes considering the OE and their performance over OE and its replicate experiment called IE. The following are the designs for the experiments:
%the dataset described in the subsection \ref{sec:methodology}:

\begin{comment}
\begin{itemize}
\item  Approach from OE without AS (Approach 1 w/o AS): only use data from the Original Experimental benchmark dataset clinical P300 dataset (OE).
\item Approach  from OE with AS (Approach 1 w/ AS): use data from the Original Experimental benchmark dataset clinical P300 dataset (OE) after applying active sampling. 
\item Approach from OE and IE without AS (Approach 2 w/o AS):  use data from Original Experimental benchmark dataset clinical P300 dataset (OE) combined with the Multi-centre benchmark clinical P300 dataset (IE).
\item Approach from OE and IE with AS (Approach 2 w/ AS): use data from Original Experimental benchmark dataset clinical P300 dataset (OE) combined with the Multi-centre benchmark clinical P300 dataset (IE) after applying active sampling. 
\end{itemize}
\end{comment}

\begin{itemize}
\item  Approach from Original experimental benchmark dataset clinical P300 dataset (OE):
\begin{itemize}
\item OE without Active Sampling (OE w/o AS): use the whole data from the Original Experimental benchmark dataset clinical P300 dataset (OE). 
\item OE with Active Sampling (OE w/ AS): use data from the Original Experimental benchmark dataset clinical P300 dataset (OE) after applying Active Sampling (AS).
\end{itemize}

\item  Approach from Original experimental benchmark dataset clinical P300 dataset (OE) combined with Multi-centre benchmark clinical P300 dataset (ME):

\begin{itemize}
\item OE+ME without Active Sampling (OE+ME w/o AS): use the whole data from the Original Experimental benchmark dataset clinical P300 dataset (OE) combined with the Multi-centre benchmark clinical P300 dataset (ME). 
\item OE+ME with Active Sampling (OE+ME w/ AS): use data from the Original Experimental benchmark dataset clinical P300 dataset (OE) combined with the Multi-centre benchmark clinical P300 dataset (ME) after applying Active Sampling (AS).
\end{itemize}

\end{itemize}

\section{Results} \label{results}
We evaluated Active Sampling (AS) with Dense PDS on Deep4Net, aiming to reduce computational cost while maintaining or improving classification accuracy. Our deep neural network, trained and tested in Google Colab using Python, underwent a cross-validation process. We experimented with different Sampling Factor (SF) sizes to find the most effective one. This evaluation included cross-validation with distinct training, validation, and testing phases across three subject strategies: dependent, independent, and adaptive. In the adaptive approach, we varied the percentages of data split from the target subject for model fine-tuning, seeking the optimal value for comparison.

\subsection{Samples number selection ( Sampling Factor)}
%In order to determine the best samples number for a Dense PSD, a first test was conducted by varying this value from 500 to 1200 samples and reporting only the best test score for subject independent training with its respectively sample size. 
To determine the best sample number, we use Active Sampling with the Dense PDS to calculate the best classification accuracy for subject-independent, varying the sampling number from 500 to 1200.
%In order to determine the best samples number for a Dense PSD, we best test score for subject independent training with its respectively sample size.  first test was conducted by varying this value from 500 to 1200 samples and reporting only the best test score for subject independent training with its respectively sample size. 

We summarize the process for both datasets using a Dense PDS:

\begin{enumerate}
    \item Calculate a subject-independent classification using a Leave-
One-Subject-Out Cross-Validation (LOSOCV) for all subjects and both datasets (OE and ME) using different sample sizes.
    \item The sample size increased from 500  to 1200 in steps of 100 and studied their performance improvements.
    \item The best subject-independent classification accuracy test and its corresponding sample size were chosen. 
    \item Create a histogram to perform the distribution of the best sample size.
\end{enumerate}

Table \ref{tab:best_sample} summarizes the results of selecting the best subject-independent classification accuracy with its sampled number. To facilitate the analysis of data from both datasets (OE and ME), we listed the number of subjects in a correlative manner.
The highest result was achieved in Subject 10, with a precision of 99.67\% to a sample size of 1200. Besides, subjects with higher values than 90\% were subjects 10,12,13,14,15 and 18 belonging to the ME dataset. Performance for the subjects of the OE dataset was positive but did not surpass 90\% classification accuracy, as subjects' have critical medical conditions. In general, 1200 samples proved to be the optimal choice.
\begin{table}[h]
\scriptsize
\centering
\caption{Selection of appropriate sample factor}
\begin{tabular}[b]{c c c c}
\toprule
\multirow{2}{*}{\textbf{Dataset}} & \multirow{2}{*}{\textbf{Subjects}} & \multirow{2}{*}{\textbf{\begin{tabular}[c]{@{}c@{}}Best Sub.Indep \\ test score\end{tabular}}} & \multirow{2}{*}{\textbf{Best sample number}} \\
 & \multicolumn{1}{c}{} &  &  \\ \midrule
\multirow{9}{*}{\textbf{OE}}  
  & S01 & 79,67\% & 1200 \\
  & S02 & 86,50\% & 1200 \\
  & S03 & 75,67\% & 1200 \\
  & S04  & 83,64\% & 1100 \\
  & S06  & 84,33\% & 1200 \\
  & S07  & 86,89\% & 900 \\
  & S08  & 73,27\% & 1100 \\
  & S09  & 73,83\% & 1200 \\
 \midrule
\multirow{8}{*}{\textbf{ME}} 
   & S10 & 99,67\% & 1200 \\
  & S11 & 84,22\% & 900 \\
  & S12 & 90,83\% & 1200 \\
  & S13 & 93,33\% & 900 \\
  & S14 & 94,50\% & 1200 \\
  & S15 & 97,09\% & 1100 \\
  & S16 & 80,80\% & 1000 \\
  & S17  & 85,83\% & 1200 \\
  & S18  & 92,50\% & 1200 \\
 \bottomrule
 & \multicolumn{1}{c}{} & \textbf{Max.freq sampled} & 1200\\
\cline{3-4} 
\end{tabular}
\label{tab:best_sample}
\end{table}

\begin{comment}
% Commentary on the reduction of pages
To analyze the Best sample number for each subject, we elaborate a histogram, as we can see in \ref{fig:best_sample_barplot}. 
It outlines that 1200 samples are the best choice since they reached 10 counts. The total of counts adds up to 17, representing the total number of subjects for two datasets. 

\begin{figure}[h]
  \centering
  \includesvg[width=0.4\textwidth]{Figures/3. Best sample barplot.svg}
  \caption{Sample size histogram with best subject independent test accuracy across subjects belonging to OE + ME}
  \label{fig:best_sample_barplot}
\end{figure}
\end{comment}

\begin{figure}[h]
  \centering  \includegraphics[width=0.36\textwidth]{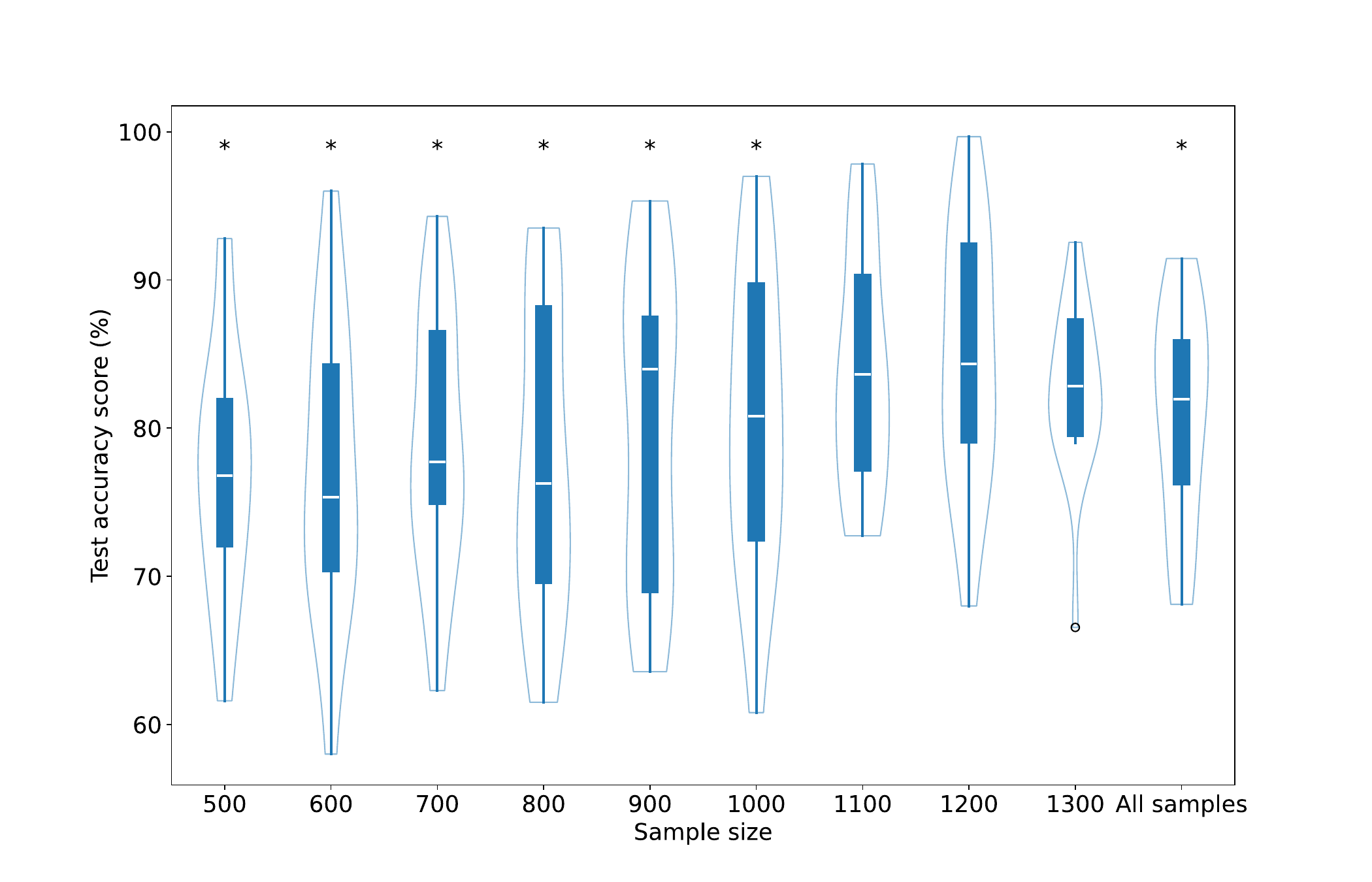}
  \caption{Violinplot classification test accuracy distributions with different sample sizes from OE + ME. The statistical test using the Wilcoxon signed-rank test with p-values (p $<$ 0.05) depicted by ($*$) means the statistical difference between the 1200 sample size and each of the other sample sizes.}
 
% with the Wilcoxon test (p $<$ 0.05).} %Also, 1100 samples results are statistically different than all other cases except 1200 and 1300.
  \label{fig:best_sample_boxplot}
\end{figure}

In \ref{fig:best_sample_boxplot}, the violinplot displays the classification test accuracy distributions for each sample size for all 17 subjects. Hence, the violin plot for the 1200 sample size reported the highest median and quartiles, while the 800 sample size reported the lowest median and the lowest first quartile. Therefore, the accuracy values reported a non-normal distribution. Thus, a non-parametric Wilcoxon signed-rank test was performed to analyze the median of differences between the 1200 sample size and the other sample sizes. The 1200 sample size reported p-values (p $<$ 0.05) with all sample sizes except the 1100 and 1300 sample sizes.  It is important to note that as the sample size increases from 500 to 1200, the third quartile also increases. However, for the 1300 sample size, the third quartile decreases, indicating that the 1200 sample size is optimal since it maintains high classification accuracy with fewer samples than the 1300 sample size or without sampling. After this thorough analysis, we selected the 1200 sample size.

%****Furthemore, 1300 samples reported a decrease in the median and quartiles.****
%48

\subsection{Sampling Factor effect over dataset}

\begin{figure}[htpb!]
    \centering
    \begin{subfigure}{0.18\textwidth}
        \includegraphics[width=\textwidth]{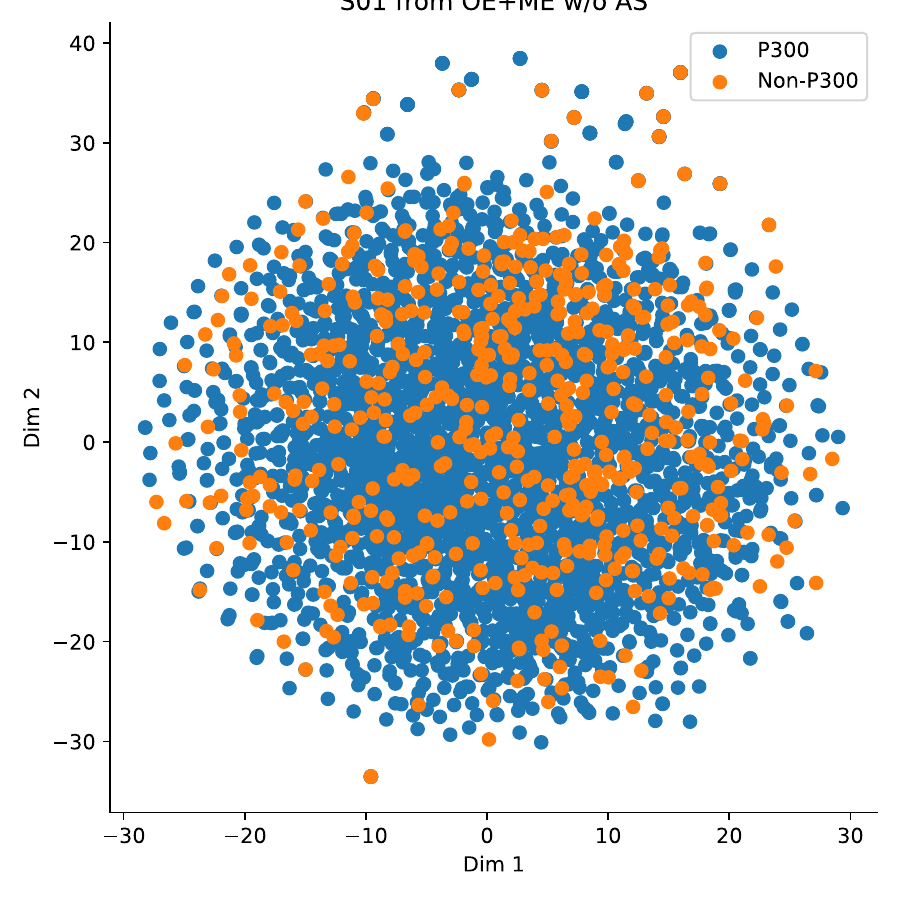}
    \end{subfigure}
    \hfill
    \begin{subfigure}{0.18\textwidth}
        \includegraphics[width=\textwidth]{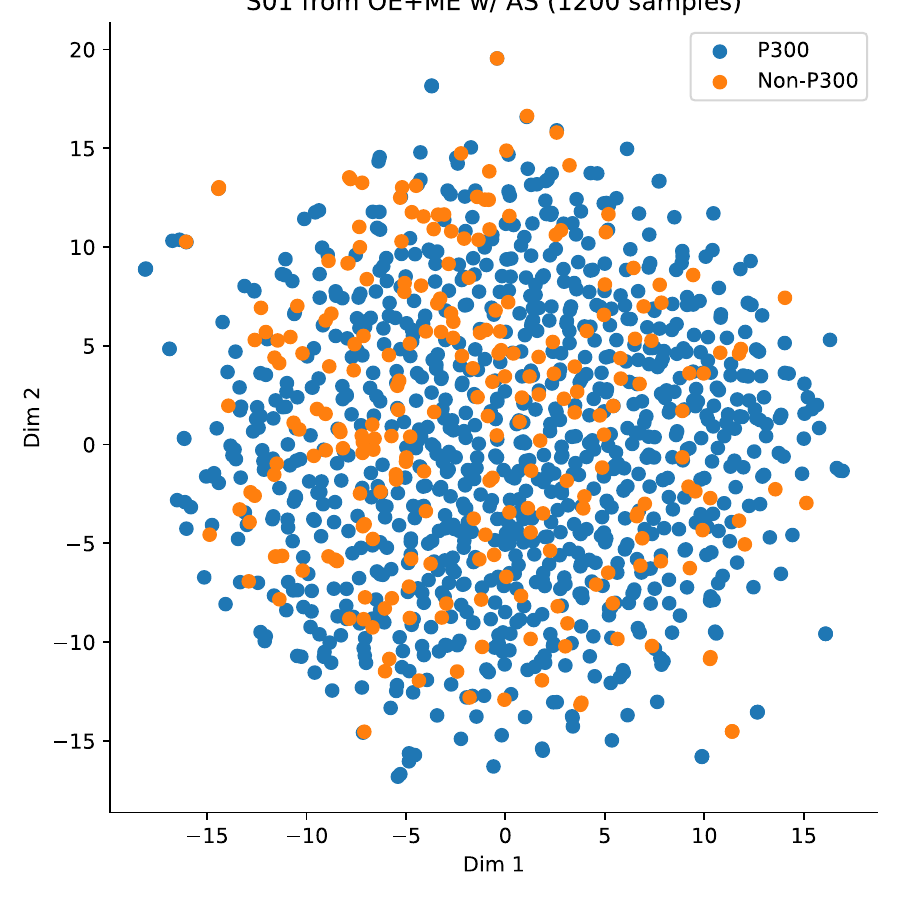}
    \end{subfigure}
    %\hfill   %%%%    
\caption{tSNE visualization of 2D feature space for S01 of OE+ME w/AS (1200 samples) and w/o AS}   
\label{pics:TSE}
\end{figure}

To create a comprehensive visualization comparison of the performance of Active Sampling (AS) using a Dense PDS over the data from each subject and dataset, we performed a t-distributed stochastic neighbour embedding (t-SNE) plot. The input data are all P300 and non-P300 samples after being pre-processed according to \ref{Data_pre_processing}  (P300-pre-processing). The aim is to compare the impact on the data before and after performing Active Sampling (AS). Thus, the experimental results are presented in Fig.\ref{pics:TSE}, where subject 01 was chosen for complete visualization of the AS effect. More results can be found in Section S.I of the supplementary material. Hence, the first row of Fig \ref{pics:TSE} depicts the t-SNE of samples before applying AS, and the second shows a t-SNE of samples after using AS. Furthermore, orange and blue dots represent the P300 and non-P300 samples, respectively. For all figures, the Active Sampling (AS) demonstrated data reduction in 1200 samples, increasing the sampling diversity, thus eliminating redundant data for better Stochastic Gradient Descent (SGD) convergence and classification accuracy. For instance, subject 01, after applying Active Sampling (AS), demonstrated a profound reduction of non-P300 samples due to their high number and the decrease of P300 samples. It leads to maintaining the data structure and precise decision boundary.
%0.20
\subsection{Classification Accuracy from OE}
\begin{figure}[h]
     \centering
     \begin{subfigure}[b]{0.24\textwidth}
         \centering
         \includegraphics[width=\textwidth]{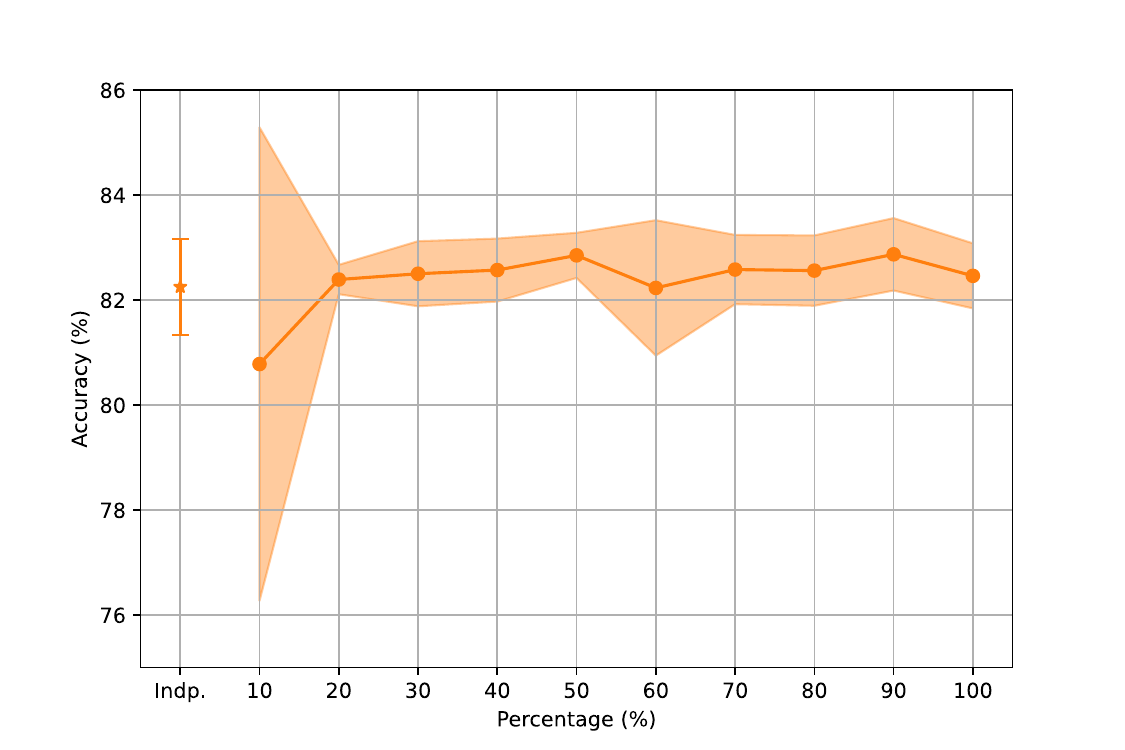}
         \caption{OE w/o AS}
         \label{fig:average_hoffman_raw}
     \end{subfigure}
     \hfill
     \begin{subfigure}[b]{0.24\textwidth}
         \centering
         \includegraphics[width=\textwidth]{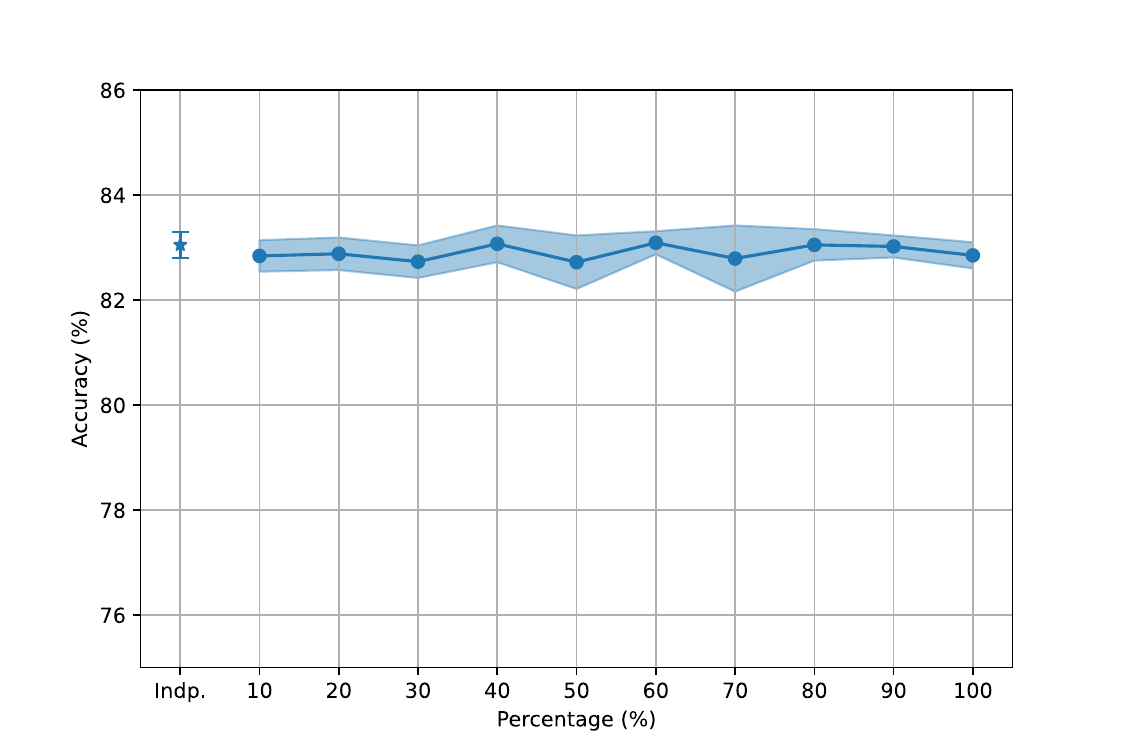}
         \caption{OE w/ AS (1200 samples)}
         \label{fig:average_hoffman_sampling}
     \end{subfigure}
     
     \begin{subfigure}[b]{0.24\textwidth}
         \centering
         \includegraphics[width=\textwidth]{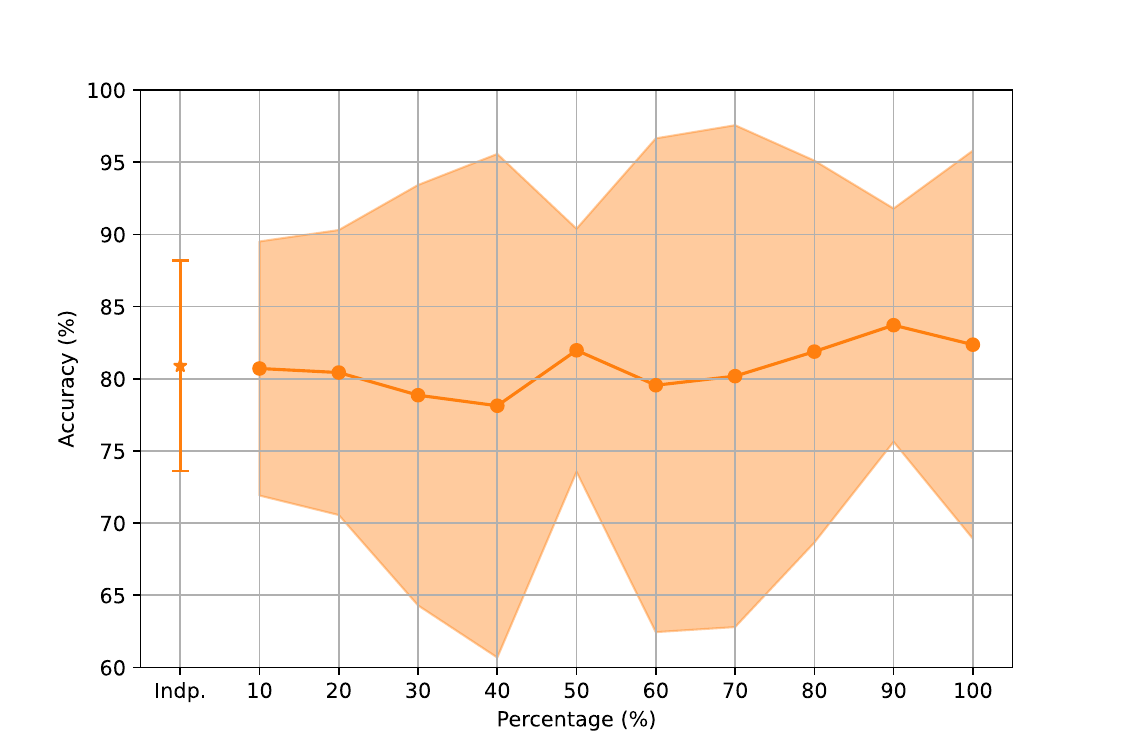}
         \caption{OE+ME w/o AS}
         \label{fig:average_multicenter_raw}
     \end{subfigure}
     \begin{subfigure}[b]{0.24\textwidth}
         \centering
         \includegraphics[width=\textwidth]{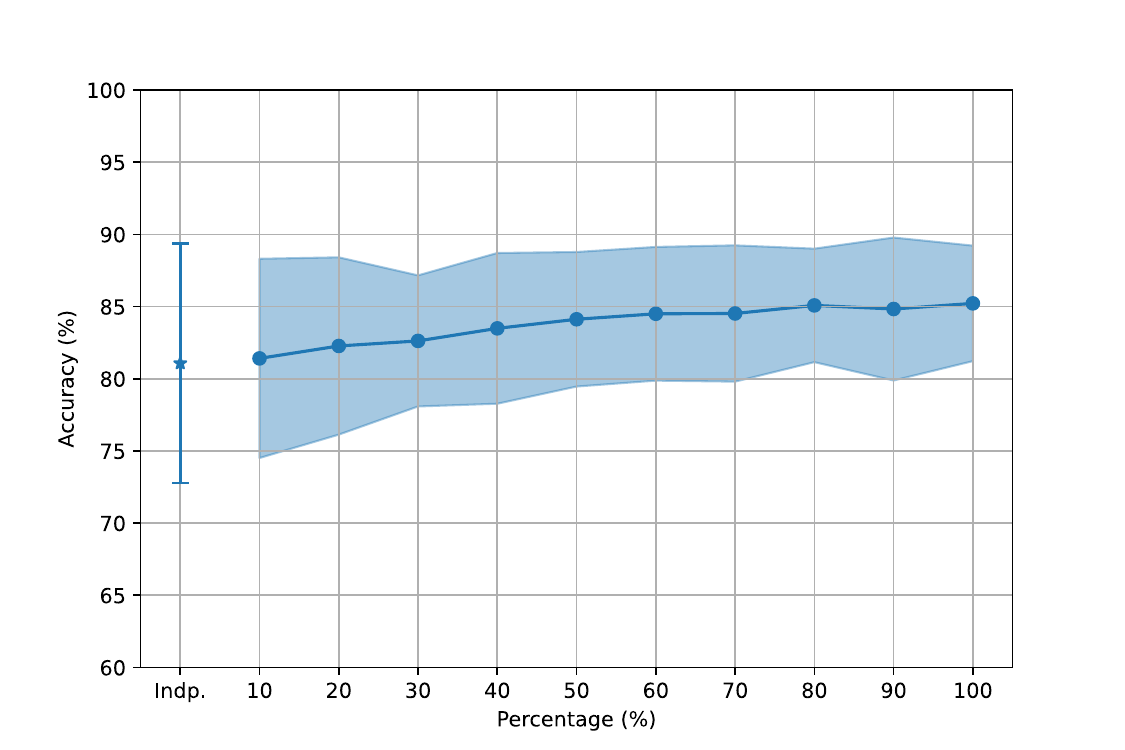}
         \caption{OE+ME w/ AS (1200 smpls.)}
         \label{fig:average_multicenter_sampling}
     \end{subfigure}
     
    \caption{Comparison of average test classification accuracy w/ AE and w/o AE across different adaptive percentages. The shadowed area represents the standard deviation.}
    \label{fig:Main_Results_subject_A1(OE)_vs_A2(OE_plus_IE)}
\end{figure}

\begin{figure}[h]
     \centering
     \begin{subfigure}[b]{0.5\textwidth}
         \centering
        \includegraphics[width=\textwidth]{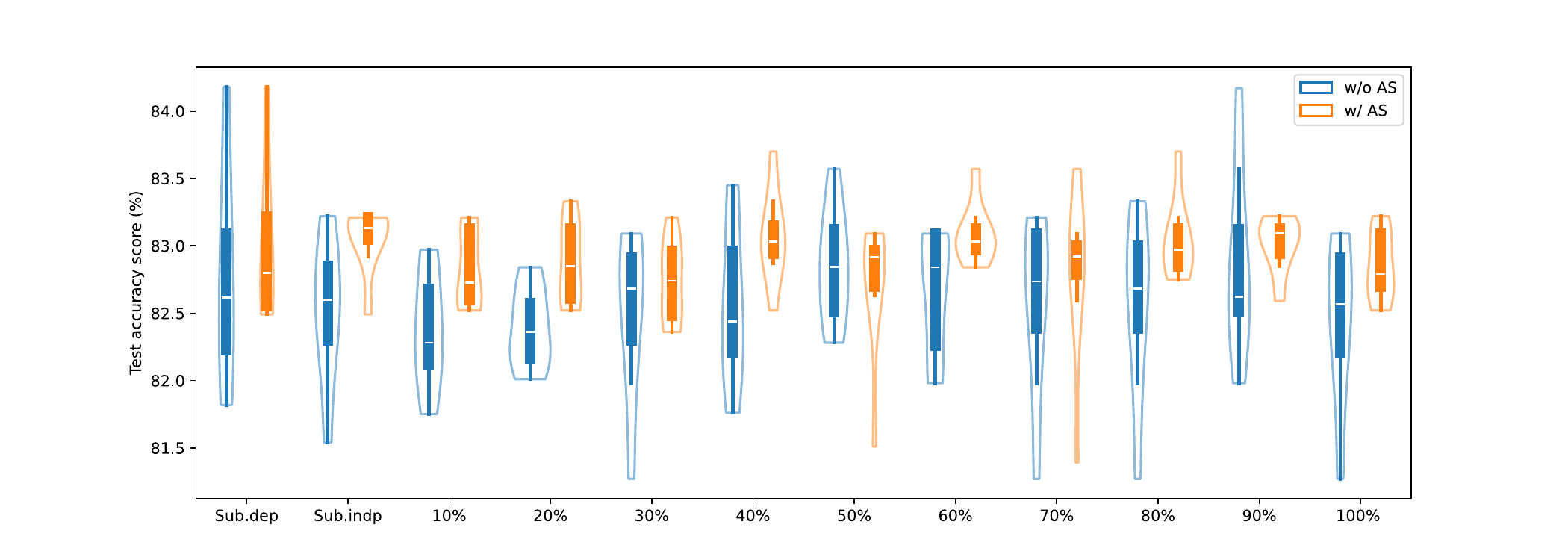}
         \caption{OE results}
         \label{fig:hoffman_accuracy}
     \end{subfigure}
     \hfill
     \begin{subfigure}[b]{0.5\textwidth}
         \centering
        \includegraphics[width=\textwidth]{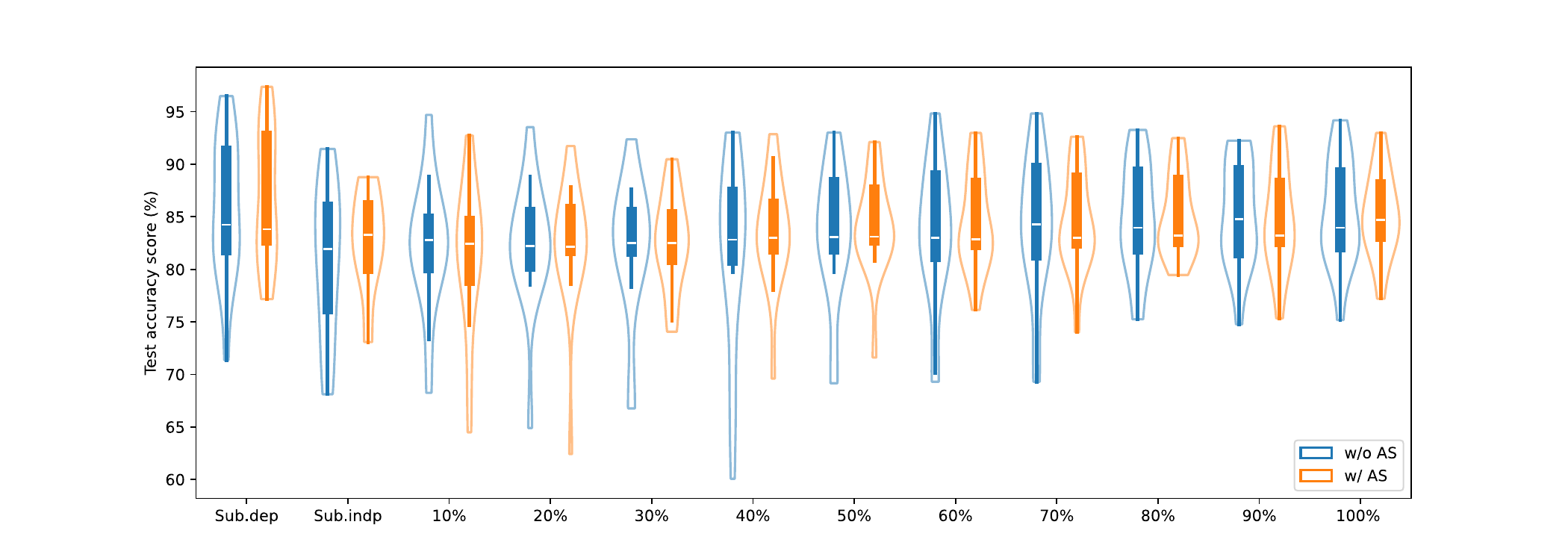}
         \caption{OE + ME results}
         \label{fig:multicenter_accuracy}
     \end{subfigure}
    \caption{Classification accuracy for w/o AS and w/ AS (1200 samples)}
    \label{Multicenter_all}
\end{figure}

\begin{figure}[h]
     \begin{subfigure}[b]{0.24\textwidth}
         \includegraphics[width=\textwidth]{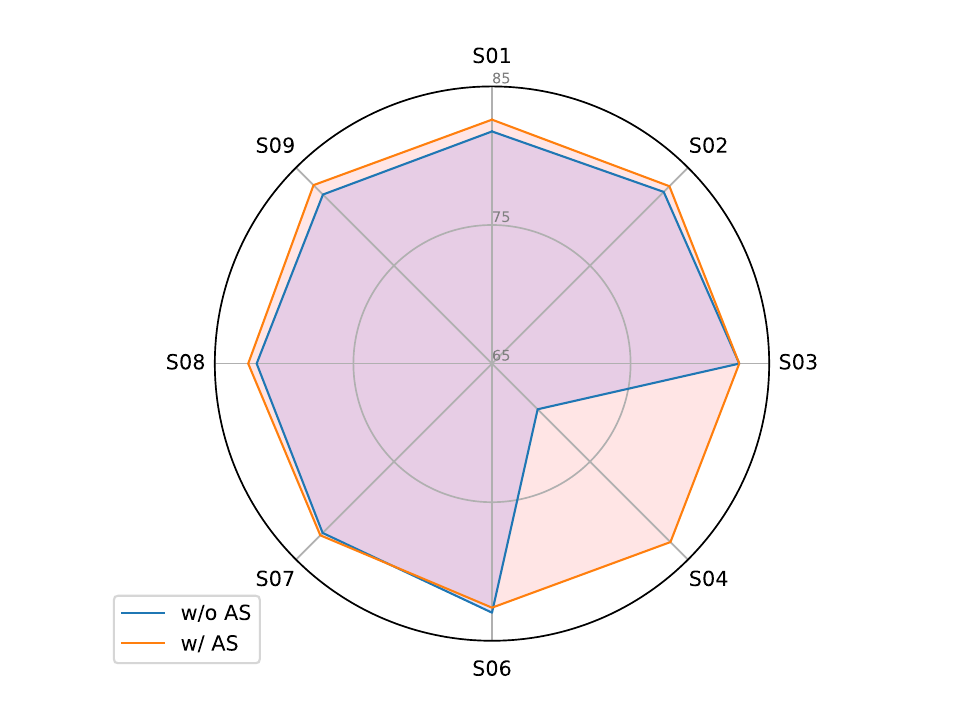}
          \caption{Best adaptive trial (10\%) for OE}
          \label{Best_adaptive_hoffman}
     \end{subfigure}
     \hfill
     \begin{subfigure}[b]{0.24\textwidth}
         \includegraphics[width=\textwidth]{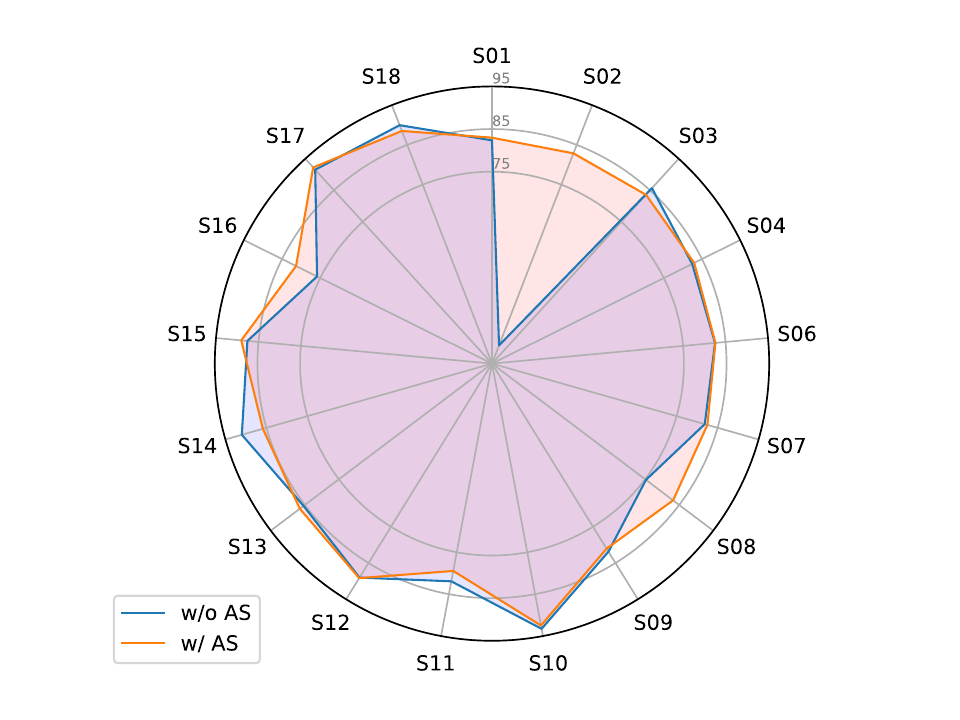}
          \caption{Best adaptive trial (80\%) for OE+ME}
          \label{Best_adaptive_multicenter}
     \end{subfigure}
    \caption{Comparison between w/o AS and w/ AS (1200 samples) for the best adaptive trial}
    \label{comparison_w_wo_sampling}
\end{figure}

This section summarizes the classification accuracy results for the OE experimental scheme, both with and without Active Sampling (AS). Tables \ref{tab:Hoffman_raw_data} and \ref{tab:Hoffman_sampling} present the average classification accuracies for all 8 subjects using OE w/o AS and OE w/ AS, respectively. Our findings indicate that OE w/ AS consistently surpasses OE w/o AS in subject-dependent, subject-independent, and subject-adaptive classifications, also reducing the standard deviation. Figure \ref{fig:Main_Results_subject_A1(OE)_vs_A2(OE_plus_IE)} illustrates these performance differences and variance among subjects.
Figure \ref{fig:average_hoffman_raw} and Figure \ref{fig:average_hoffman_sampling} highlight the impact of OE w/o AS and OE w/ AS. The use of AS yields a lower standard deviation across all adaptation rates and maintains an average classification accuracy of approximately 83\%, outperforming OE w/o AS. This demonstrates AS's effectiveness in reducing computational time and standard deviation.
Violin plots in Figure \ref{fig:hoffman_accuracy} offer a robust statistical analysis, showing a higher median and a smaller range for OE w/ AS compared to OE w/o AS across all classification schemes. This indicates a reduction in the interquartile range and a compression of the maximum and minimum values for OE w/ AS.
%$-SP$ ****
For the best case in the adaptive scheme with a 10\% adaptation rate, using 1200 samples typically improves classification accuracy by 0.1\% to 13\% for subjects 1,2,4,7,8 and 9, as shown in Figure \ref{Best_adaptive_hoffman}. Subject 4 notably exhibited an improvement of about 13\% with the use of AS.

\begin{table*}[t]
\tiny	
\caption{Classification accuracy of OE w/o AS}
\begin{tabular}{cllcccccccccccc}
\hline
\multicolumn{1}{c}{\multirow{2}{*}{\textbf{Dataset}}} & \multicolumn{1}{c}{\multirow{2}{*}{\textbf{Diagnosis}}} & \multicolumn{1}{c}{\multirow{2}{*}{\textbf{Subjects}}} & \multirow{2}{*}{\textbf{Subject-dependent}} & \multirow{2}{*}{\textbf{Subject-independent}} & \multicolumn{10}{c}{\textbf{Subject-adaptive}}  \\ \cline{6-15} 

\multicolumn{1}{c}{}  & \multicolumn{1}{c}{}  & \multicolumn{1}{c}{} &  &  & \multicolumn{1}{c}{\textbf{10\%}} & \multicolumn{1}{c}{\textbf{20\%}} & \multicolumn{1}{c}{\textbf{30\%}} & \multicolumn{1}{c}{\textbf{40\%}} & \multicolumn{1}{c}{\textbf{50\%}} & \multicolumn{1}{c}{\textbf{60\%}} & \multicolumn{1}{c}{\textbf{70\%}} & \multicolumn{1}{c}{\textbf{80\%}} & \multicolumn{1}{c}{\textbf{90\%}} & \textbf{100\%} \\ \hline

\multicolumn{1}{c}{\multirow{8}{*}{\textbf{OE}}} & \textbf{CP} & \textbf{S01} & 80,41\%  & 80,37\% & \multicolumn{1}{c}{81,75\%} & \multicolumn{1}{c}{82,12\%}  & \multicolumn{1}{c}{81,27\%} & \multicolumn{1}{c}{82,36\%}       & \multicolumn{1}{c}{82,85\%}       & \multicolumn{1}{c}{82,00\%} & \multicolumn{1}{c}{81,27\%} & \multicolumn{1}{c}{81,27\%}  & \multicolumn{1}{c}{\textbf{82,97\%}}& 81,27\%        \\ \cline{2-15}  

\multicolumn{1}{c}{}  & \textbf{MS} & \textbf{S02} & 81,82\% & 82,38\%  & 
\multicolumn{1}{c}{82,52\%} & \multicolumn{1}{c}{82,52\%} & \multicolumn{1}{c}{82,52\%}  & \multicolumn{1}{c}{82,52\%} & \multicolumn{1}{c}{82,52\%} & \multicolumn{1}{c}{82,52\%}   & \multicolumn{1}{c}{82,52\%}  & \multicolumn{1}{c}{82,52\%}  & \multicolumn{1}{c}{82,52\%} & 82,52\%  \\ \cline{2-15}

\multicolumn{1}{c}{} & \textbf{Late stage ALS} & \textbf{S03}& 83,09\% & 82,97\% & \multicolumn{1}{c}{82,84\%} & \multicolumn{1}{c}{82,84\%}  & \multicolumn{1}{c}{82,84\%}  & \multicolumn{1}{c}{82,84\%}  & \multicolumn{1}{c}{82,84\%}  & \multicolumn{1}{c}{82,84\%} & \multicolumn{1}{c}{82,84\%}  & \multicolumn{1}{c}{82,84\%} & \multicolumn{1}{c}{82,72\%}  & \textbf{83,09\%}  \\ \cline{2-15} 

\multicolumn{1}{c}{}  & \textbf{TBI \& SCI level C4} & \textbf{S04} & 82,61\%  & 81,54\%  & \multicolumn{1}{c}{69,66\%} & \multicolumn{1}{c}{82,01\%} & \multicolumn{1}{c}{83,09\%}   & \multicolumn{1}{c}{83,45\%}  & \multicolumn{1}{c}{83,57\%} & \multicolumn{1}{c}{83,09\%}  & \multicolumn{1}{c}{83,09\%}  & \multicolumn{1}{c}{83,33\%}  & \multicolumn{1}{c}{\textbf{84,17\%}} & 82,85\%  \\ \cline{2-15}

\multicolumn{1}{c}{} & \textbf{Healthy} & \textbf{S06} & 84,18\%  & 82,73\%& \multicolumn{1}{c}{82,97\%}& \multicolumn{1}{c}{82,25\%} & \multicolumn{1}{c}{82,85\%}  & \multicolumn{1}{c}{83,33\%}  & \multicolumn{1}{c}{83,21\%}& \multicolumn{1}{c}{83,09\%} & \multicolumn{1}{c}{83,21\%}  & \multicolumn{1}{c}{82,97\%}  & \multicolumn{1}{c}{\textbf{83,57\%}}  & 82,61\% \\ \cline{2-15}

\multicolumn{1}{c}{}  & \textbf{Healthy} & \textbf{S07}  & 85,20\%  & 82,21\%  & \multicolumn{1}{c}{82,28\%} & \multicolumn{1}{c}{82,17\%} & \multicolumn{1}{c}{82,40\%} & \multicolumn{1}{c}{82,28\%}  & \multicolumn{1}{c}{82,28\%}  & \multicolumn{1}{c}{79,25\%}  & \multicolumn{1}{c}{\textbf{82,63\%}} & \multicolumn{1}{c}{82,52\%}  & \multicolumn{1}{c}{82,52\%} & 82,28\%  \\ \cline{2-15} 

\multicolumn{1}{c}{} & \textbf{Healthy}& \textbf{S08} & 82,22\%  & \textbf{82,60\%}        & \multicolumn{1}{c}{81,98\%}& \multicolumn{1}{c}{82,47\%} & \multicolumn{1}{c}{81,98\%}       & \multicolumn{1}{c}{81,98\%} & \multicolumn{1}{c}{82,47\%} & \multicolumn{1}{c}{81,98\%} & \multicolumn{1}{c}{81,98\%} & \multicolumn{1}{c}{81,98\%} & \multicolumn{1}{c}{81,98\%} & 81,98\% \\ \cline{2-15}

\multicolumn{1}{c}{}  & \textbf{Healthy} & \textbf{S09} & 71,74\%  & \textbf{83,22\%}  & \multicolumn{1}{c}{82,25\%}  & \multicolumn{1}{c}{82,73\%}  & \multicolumn{1}{c}{83,09\%}   & \multicolumn{1}{c}{81,76\%}  & \multicolumn{1}{c}{83,09\%}  & \multicolumn{1}{c}{83,09\%} & \multicolumn{1}{c}{83,09\%}  & \multicolumn{1}{c}{83,09\%}       & \multicolumn{1}{c}{82,49\%} & 83,09\%    \\ \hline

\multicolumn{1}{l}{} &  & \multicolumn{1}{c}{\textbf{Mean}}  & 81,41\% & 82,25\% & \multicolumn{1}{c}{80,78\%}  & \multicolumn{1}{c}{82,39\%}  & \multicolumn{1}{c}{82,50\%}  & \multicolumn{1}{c}{82,57\%}  & \multicolumn{1}{c}{82,85\%}    & \multicolumn{1}{c}{82,23\%}  & \multicolumn{1}{c}{82,58\%}& \multicolumn{1}{c}{82,56\%} & \multicolumn{1}{c}{\textbf{82,87\%}}& 82,46\% \\ \cline{3-15} 

\multicolumn{1}{l}{}  &  & \multicolumn{1}{c}{\textbf{Std.dev}} & 4,17\% & 0,92\% & \multicolumn{1}{c}{4,51\%}  & \multicolumn{1}{c}{\textbf{0,30\%}} & \multicolumn{1}{c}{0,62\%}  & \multicolumn{1}{c}{0,60\%}   & \multicolumn{1}{c}{0,43\%}  & \multicolumn{1}{c}{1,29\%}  & \multicolumn{1}{c}{0,66\%}   & \multicolumn{1}{c}{0,67\%}  & \multicolumn{1}{c}{0,70\%} & 0,62\%   \\ \cline{3-15} 
\end{tabular}
\label{tab:Hoffman_raw_data}
\end{table*}

% Table: 1200 sampled Hoffman
\begin{table*}[t]
\tiny
\caption{Classification accuracy of OE w/ AS (1200 samples)}
\begin{tabular}{cllcccccccccccc}
\hline
\multicolumn{1}{c}{\multirow{2}{*}{\textbf{Dataset}}} & \multicolumn{1}{c}{\multirow{2}{*}{\textbf{Diagnosis}}} & \multicolumn{1}{c}{\multirow{2}{*}{\textbf{Subjects}}} & \multirow{2}{*}{\textbf{Subject-dependent}} & \multirow{2}{*}{\textbf{Subject-independent}} & \multicolumn{10}{c}{\textbf{Subject-adaptive}}                                                                                      \\ \cline{6-15} \multicolumn{1}{c}{} & \multicolumn{1}{c}{}                                    & \multicolumn{1}{c}{}  &   &  & \multicolumn{1}{c}{\textbf{10\%}} & \multicolumn{1}{c}{\textbf{20\%}} & \multicolumn{1}{c}{\textbf{30\%}} & \multicolumn{1}{c}{\textbf{40\%}} & \multicolumn{1}{c}{\textbf{50\%}} & \multicolumn{1}{c}{\textbf{60\%}} & \multicolumn{1}{c}{\textbf{70\%}} & \multicolumn{1}{c}{\textbf{80\%}} & \multicolumn{1}{c}{\textbf{90\%}} & \textbf{100\%} \\ \hline

\multicolumn{1}{c}{\multirow{8}{*}{\textbf{OE}}} & \textbf{CP}                & \textbf{S01}                                    & \textbf{84,18\%}                                    & 82,49\%                         & \multicolumn{1}{c}{82,60\%}       & \multicolumn{1}{c}{82,73\%}       & \multicolumn{1}{c}{82,36\%}       & \multicolumn{1}{c}{83,09\%}       & \multicolumn{1}{c}{81,51\%}       & \multicolumn{1}{c}{83,21\%}       & \multicolumn{1}{c}{81,39\%}       & \multicolumn{1}{c}{83,70\%}       & \multicolumn{1}{c}{83,09\%}       & 82,85\%        

\\ \cline{2-15} 
\multicolumn{1}{c}{}& \textbf{MS}                                              & \textbf{S02}                                    & 80,30\%                                    & 82,92\%                             & \multicolumn{1}{c}{83,10\%}       & \multicolumn{1}{c}{83,10\%}       & \multicolumn{1}{c}{82,87\%}       & \multicolumn{1}{c}{82,87\%}       & \multicolumn{1}{c}{82,87\%}       & \multicolumn{1}{c}{83,10\%}       & \multicolumn{1}{c}{82,87\%}       & \multicolumn{1}{c}{83,10\%}       & \multicolumn{1}{c}{\textbf{83,22\%}}       & 83,22\%       

\\ \cline{2-15} 
\multicolumn{1}{c}{}                                           & \textbf{Late stage ALS}                             & \textbf{S03}                                    & \textbf{88,73\%}                                    & 83,21\%                                       & \multicolumn{1}{c}{82,84\%}       & \multicolumn{1}{c}{82,97\%}       & \multicolumn{1}{c}{82,97\%}       & \multicolumn{1}{c}{83,70\%}       & \multicolumn{1}{c}{82,72\%}       & \multicolumn{1}{c}{82,97\%}       & \multicolumn{1}{c}{83,09\%}       & \multicolumn{1}{c}{82,84\%}       & \multicolumn{1}{c}{83,09\%}       & 83,09\%        

\\ \cline{2-15} 
\multicolumn{1}{c}{}                                           & \textbf{TBI \& SCI level C4}                   & \textbf{S04}                                    & 82,73\%                                    & 83,21\%                                       & \multicolumn{1}{c}{83,21\%}       & \multicolumn{1}{c}{83,21\%}       & \multicolumn{1}{c}{82,37\%}       & \multicolumn{1}{c}{82,97\%}       & \multicolumn{1}{c}{83,09\%}       & \multicolumn{1}{c}{82,97\%}       & \multicolumn{1}{c}{82,97\%}       & \multicolumn{1}{c}{82,85\%}       & \multicolumn{1}{c}{82,85\%}       & 82,73\%        

\\ \cline{2-15} 
\multicolumn{1}{c}{}                                           & \textbf{Healthy}                                         & \textbf{S06}                                    & 82,49\%                                    & 83,09\%                                       & \multicolumn{1}{c}{82,61\%}       & \multicolumn{1}{c}{82,61\%}       & \multicolumn{1}{c}{82,61\%}       & \multicolumn{1}{c}{83,09\%}       & \multicolumn{1}{c}{82,97\%}       & \multicolumn{1}{c}{82,97\%}       & \multicolumn{1}{c}{82,97\%}       & \multicolumn{1}{c}{82,97\%}       & \multicolumn{1}{c}{82,97\%}       & 82,73\%        

\\ \cline{2-15} 
\multicolumn{1}{c}{}                                           & \textbf{Healthy}                                         & \textbf{S07}                                    & 82,87\%                                    & 83,10\%                                       & \multicolumn{1}{c}{82,52\%}       & \multicolumn{1}{c}{82,52\%}       & \multicolumn{1}{c}{82,52\%}       & \multicolumn{1}{c}{82,52\%}       & \multicolumn{1}{c}{82,63\%}       & \multicolumn{1}{c}{83,57\%}       & \multicolumn{1}{c}{83,57\%}       & \multicolumn{1}{c}{82,75\%}       & \multicolumn{1}{c}{83,10\%}       & 82,52\%        

\\ \cline{2-15} 
\multicolumn{1}{c}{}                                           & \textbf{Healthy}                                         & \textbf{S08}                                    & \textbf{83,33\%}                                    & 83,21\%                                       & \multicolumn{1}{c}{82,59\%}       & \multicolumn{1}{c}{82,59\%}       & \multicolumn{1}{c}{82,96\%}       & \multicolumn{1}{c}{82,96\%}       & \multicolumn{1}{c}{82,96\%}       & \multicolumn{1}{c}{82,84\%}       & \multicolumn{1}{c}{82,59\%}       & \multicolumn{1}{c}{83,21\%}       & \multicolumn{1}{c}{82,59\%}       & 82,59\%        

\\ \cline{2-15} 
\multicolumn{1}{c}{}                                           & \textbf{Healthy}                                         & \textbf{S09}                                    & 82,49\%                                    & 83,16\%                                       & \multicolumn{1}{c}{83,21\%}       & \multicolumn{1}{c}{83,33\%}       & \multicolumn{1}{c}{83,21\%}       & \multicolumn{1}{c}{83,33\%}       & \multicolumn{1}{c}{82,97\%}       & \multicolumn{1}{c}{83,09\%}       & \multicolumn{1}{c}{82,85\%}       & \multicolumn{1}{c}{82,97\%}       & \multicolumn{1}{c}{83,21\%}       & 83,09\%        

\\ \hline
\multicolumn{1}{l}{}                                             &                                                          & \multicolumn{1}{c}{\textbf{Mean}}                      & \textbf{83,39\%}                                    & 83,05\%                                       & \multicolumn{1}{c}{82,84\%}       & \multicolumn{1}{c}{82,88\%}       & \multicolumn{1}{c}{82,73\%}       & \multicolumn{1}{c}{83,07\%}       & \multicolumn{1}{c}{82,72\%}       & \multicolumn{1}{c}{83,09\%}       & \multicolumn{1}{c}{82,79\%}       & \multicolumn{1}{c}{83,05\%}       & \multicolumn{1}{c}{83,02\%}       & 82,85\%        
\\ \cline{3-15} 
\multicolumn{1}{l}{}                                             &                                                          & \multicolumn{1}{c}{\textbf{Std.dev}}                   & 2,42\%                                     & 0,25\%                                        & \multicolumn{1}{c}{0,30\%}        & \multicolumn{1}{c}{0,31\%}        & \multicolumn{1}{c}{0,31\%}        & \multicolumn{1}{c}{0,35\%}        & \multicolumn{1}{c}{0,51\%}        & \multicolumn{1}{c}{0,22\%}        & \multicolumn{1}{c}{0,63\%}        & \multicolumn{1}{c}{0,30\%}        & \multicolumn{1}{c}{\textbf{0,21\%}}        & 0,25\%         \\ \cline{3-15} 
\end{tabular}
\label{tab:Hoffman_sampling}
\end{table*}

% Classification accuracy of original experiment for raw data and sampling 1200 data

% Improvement of original center over independent of raw data for sampling and raw data itself

% Best Adpative for Original experiment

%\subsection{Classification Accuracy of Extendend multi-center experiment}
%\subsection{Classification Accuracy of Approach from OE and IE}

\subsection{Classification Accuracy from OE and ME}
This section showcases the classification accuracy outcomes from the experimental approach that combines data from the Original Experiment (OE) and the Multi-centre Experiment (ME). It highlights how Active Sampling (AS) excels in scaling both accuracy and training duration when handling two different datasets.
%This section presents the results of classification accuracy for the experimental scheme from OE+ME either for OE+ME w/o AS and OE+ME w/ AS.
% Table
Tables \ref{tab:Multicenter_raw_data} and \ref{tab:Multicenter_sampling} tabulate the average classification accuracy to compare the performances across all 17 subjects using OE+ME w/o AS and OE+ME w/ AS, respectively. Our results reported that the average classification accuracy of OE+ME w/ AS outperformed OE+ME w/o AS for subject-independent and subject-adaptive except for subject-dependent. Therefore, subject-adaptive reduced the standard deviation after applying the AS for all adaptation rates. Our results reported that the average classification accuracy for a subject adaptive with 40\%  of the adaptation rate in OE+ME w/o AS outperformed OE+ME w/AS with 5.36\% and a standard deviation reduction of 12.22\%.

% General Figure
We created Figure \ref{fig:Main_Results_subject_A1(OE)_vs_A2(OE_plus_IE)} to display classification accuracy and variance among subjects for three classification schemes with and without AS.
Figure \ref{fig:average_multicenter_raw} and Figure \ref{fig:average_multicenter_sampling} show the OE+ME  w/o AS and OE+ME w/ AS, respectively. The OE +ME  w/ AS results reported a lower standard deviation for subject-adaptive for all adaptation rates compared to OE+ME  w/o AS using less computational time. 
The classification accuracy tendency for subject-adaptive in OE+ME w/o AS starts at 80.71\% for 10\% of the adaptation rate, decreasing until 78,13\% for 40\% of the adaptation rate after that, it reported a high pick value at 81,97\% for 50\%. Therefore, from 60\% to 90\% of the adaptation rate increased to 79,55\% and 83,71\%, respectively. Finally, it reached 82,36\% for 100\% of the adaptation rate. In summary, these results reported a fluctuating classification accuracy tendency. Conversely, the classification accuracy tendency for Subject Adaptive in OE+ME w/ AS increases to 81.41\%, until 85,22\%, except for a slight deflection in  90\% of the adaptation rate with 84,83\% of classification accuracy. The classification accuracy increases as adaptation rates increase, demonstrating that OE+ME w/ AS yields better results after applying AS for the adaptive scheme.

%  Violin_Box plot
In order to perform a more robust statistical analysis against outliers, boxplots were computed, grouping all subjects for the different classification schemes where the improvement offered by Active Sampling (AS) could be compared side by side. 
Figure \ref{fig:multicenter_accuracy} shows an increase in the median for OE+ME  w/ AS versus OE+ME w/o AS for subject-dependent and subject-independent. Otherwise, the medians are slightly similar among the whole subject-adaptive scheme as well as the range of the boxes. Although the medians did not increase after applying Active Sampling (AS), it reported similar results using only 1200 samples, reducing the computational cost. % Improve
Therefore, to assess the impact of active sampling (AS), we examine the change in classification accuracy between subject-independent and subject-adaptive approaches after implementing Active Sampling (AS) at various adaptation rates. The results can be found in Figure S2 in Section S.IV of the supplementary material.

%-%
%Best
Therefore, Figure \ref{Best_adaptive_multicenter}  provides additional information by comparing the classification accuracy results for the best adaptive trial before and after utilizing the Active Sampling (AS) method.
For 1200 samples and the best adaptation rate of 80\%, the Active Sampling (AS) increases the accuracy in a range of 0.1\% to 49\% for subjects 1,2,4,6,7,8,12,13,15,16 and 17
reducing their standard deviation as we described in the previous lines.
It is important to highlight that the classification accuracy of subject 2 undergoes a considerable improvement of about 49\%. 
%The rest of the subjects reported similar classification accuracy for OE+IE w/o AS and OE+IE w/ AS. Still, they reduced the standard deviation after applying the Active Sampling (AS) as we described in the previous lines.
 
% Table Classification accuracy of Raw data from Hoffman
\begin{table*}[]
\tiny
\caption{Classification of accuracy of OE+ME w/o AS}
%\color{blue} 
\begin{tabular}{cllcccccccccccc}
\hline
\multicolumn{1}{c}{\multirow{2}{*}{\textbf{Dataset}}} & \multicolumn{1}{c}{\multirow{2}{*}{\textbf{Diagnosis}}} & \multicolumn{1}{c}{\multirow{2}{*}{\textbf{Subjects}}} & \multicolumn{1}{c}{\multirow{2}{*}{\textbf{Subject-dependent}}} & \multirow{2}{*}{\textbf{Subject-independent}} & \multicolumn{10}{c}{\textbf{Subject-adaptive}}                                                                                                            \\ \cline{6-15} \multicolumn{1}{c}{} & \multicolumn{1}{c}{}  & \multicolumn{1}{c}{}                                   & \multicolumn{1}{c}{}    &  & \multicolumn{1}{c}{\textbf{10\%}} & \multicolumn{1}{c}{\textbf{20\%}} & \multicolumn{1}{c}{\textbf{30\%}} & \multicolumn{1}{c}{\textbf{40\%}} & \multicolumn{1}{c}{\textbf{50\%}} & \multicolumn{1}{c}{\textbf{60\%}} & \multicolumn{1}{c}{\textbf{70\%}} & \multicolumn{1}{c}{\textbf{80\%}} & \multicolumn{1}{c}{\textbf{90\%}} & \textbf{100\%} 

\\ \hline

\multicolumn{1}{c}{\multirow{8}{*}{\textbf{OE}}}          & \textbf{CP}                                              & \textbf{S01}                                   & \multicolumn{1}{c}{82,42\%}                                                         & \multicolumn{1}{c}{68,94\%}                                       & \multicolumn{1}{c}{68,24\%}       & \multicolumn{1}{c}{82,14\%}       & \multicolumn{1}{c}{81,93\%}       & \multicolumn{1}{c}{82,67\%}       & \multicolumn{1}{c}{\textbf{83,10\%}}       & \multicolumn{1}{c}{69,30\%}       & \multicolumn{1}{c}{69,30\%}       & \multicolumn{1}{c}{82,35\%}       & \multicolumn{1}{c}{82,25\%}       & 82,99\%        \\ \cline{2-15} 
\multicolumn{1}{c}{}                                           & \textbf{MS}                                              & \textbf{S02}                                   & \multicolumn{1}{c}{81,83\%}                                                         & \multicolumn{1}{c}{\textbf{81,95\%}}                                       & \multicolumn{1}{c}{55,18\%}       & \multicolumn{1}{c}{64,89\%}       & \multicolumn{1}{c}{28,24\%}       & \multicolumn{1}{c}{19,63\%}       & \multicolumn{1}{c}{70,56\%}       & \multicolumn{1}{c}{81,03\%}       & \multicolumn{1}{c}{81,35\%}       & \multicolumn{1}{c}{34,57\%}       & \multicolumn{1}{c}{58,67\%}       & 33,70\%        \\ \cline{2-15} 
\multicolumn{1}{c}{}                                           & \textbf{Late stage ALS}                             & \textbf{S03}                                   & \multicolumn{1}{c}{\textbf{88,88\%}}                                                         & \multicolumn{1}{c}{71,15\%}                                       & \multicolumn{1}{c}{82,99\%}       & \multicolumn{1}{c}{49,63\%}       & \multicolumn{1}{c}{85,13\%}       & \multicolumn{1}{c}{82,78\%}       & \multicolumn{1}{c}{82,99\%}       & \multicolumn{1}{c}{82,99\%}       & \multicolumn{1}{c}{85,45\%}       & \multicolumn{1}{c}{85,67\%}       & \multicolumn{1}{c}{86,63\%}       & 84,71\%        \\ \cline{2-15} 
\multicolumn{1}{c}{}                                           & \textbf{TBI \& SCI level C4}                              & \textbf{S04}                                   & \multicolumn{1}{c}{80,00\%}                                                         & \multicolumn{1}{c}{76,19\%}                                       & \multicolumn{1}{c}{79,80\%}       & \multicolumn{1}{c}{81,56\%}       & \multicolumn{1}{c}{82,33\%}       & \multicolumn{1}{c}{84,08\%}       & \multicolumn{1}{c}{\textbf{84,30\%}}       & \multicolumn{1}{c}{83,32\%}       & \multicolumn{1}{c}{83,32\%}       & \multicolumn{1}{c}{82,44\%}       & \multicolumn{1}{c}{83,42\%}       & 83,21\%        \\ \cline{2-15} 
\multicolumn{1}{c}{}                                           & \textbf{Healthy}                                         & \textbf{S06}                                   & \multicolumn{1}{c}{\textbf{84,24\%}}                                                         & \multicolumn{1}{c}{82,36\%}                                       & \multicolumn{1}{c}{83,20\%}       & \multicolumn{1}{c}{80,18\%}       & \multicolumn{1}{c}{78,28\%}       & \multicolumn{1}{c}{82,87\%}       & \multicolumn{1}{c}{82,64\%}       & \multicolumn{1}{c}{82,42\%}       & \multicolumn{1}{c}{82,31\%}       & \multicolumn{1}{c}{82,53\%}       & \multicolumn{1}{c}{82,31\%}       & 82,42\%        \\ \cline{2-15} 
\multicolumn{1}{c}{}                                           & \textbf{Healthy}                                         & \textbf{S07}                                   & \multicolumn{1}{c}{\textbf{83,51\%}}                                                         & \multicolumn{1}{c}{76,87\%}                                       & \multicolumn{1}{c}{80,52\%}       & \multicolumn{1}{c}{81,04\%}       & \multicolumn{1}{c}{81,76\%}       & \multicolumn{1}{c}{81,24\%}       & \multicolumn{1}{c}{81,55\%}       & \multicolumn{1}{c}{80,31\%}       & \multicolumn{1}{c}{81,35\%}       & \multicolumn{1}{c}{81,87\%}       & \multicolumn{1}{c}{81,35\%}       & 81,55\%        \\ \cline{2-15} 
\multicolumn{1}{c}{}                                           & \textbf{Healthy}                                         & \textbf{S08}                                   & \multicolumn{1}{c}{82,61\%}                                                         & \multicolumn{1}{c}{68,11\%}                                       & \multicolumn{1}{c}{82,13\%}       & \multicolumn{1}{c}{82,24\%}       & \multicolumn{1}{c}{\textbf{82,67\%}}       & \multicolumn{1}{c}{60,06\%}       & \multicolumn{1}{c}{59,96\%}       & \multicolumn{1}{c}{18,19\%}       & \multicolumn{1}{c}{17,65\%}       & \multicolumn{1}{c}{75,24\%}       & \multicolumn{1}{c}{81,38\%}       & 81,27\%        \\ \cline{2-15} 
\multicolumn{1}{c}{}                                           & \textbf{Healthy}                                         & \textbf{S09}                                   & \multicolumn{1}{c}{81,14\%}                                                         & \multicolumn{1}{c}{81,13\%}                                       & \multicolumn{1}{c}{\textbf{82,54\%}}       & \multicolumn{1}{c}{82,31\%}       & \multicolumn{1}{c}{81,98\%}       & \multicolumn{1}{c}{81,87\%}       & \multicolumn{1}{c}{81,98\%}       & \multicolumn{1}{c}{82,09\%}       & \multicolumn{1}{c}{82,42\%}       & \multicolumn{1}{c}{81,87\%}       & \multicolumn{1}{c}{81,20\%}       & 82,20\%   \\ \hline

\multicolumn{1}{c}{\multirow{9}{*}{\textbf{ME}}}    & \textbf{Healthy}                                         & \textbf{S10}                                    & \multicolumn{1}{c}{\textbf{96,48\%}}                                                        & \multicolumn{1}{c}{91,45\%}                                    & \multicolumn{1}{c}{94,70\%}       & \multicolumn{1}{c}{93,53\%}       & \multicolumn{1}{c}{92,37\%}       & \multicolumn{1}{c}{93,01\%}       & \multicolumn{1}{c}{93,01\%}       & \multicolumn{1}{c}{94,83\%}       & \multicolumn{1}{c}{94,83\%}       & \multicolumn{1}{c}{93,27\%}       & \multicolumn{1}{c}{92,24\%}       & 94,18\%        \\ \cline{2-15} 
\multicolumn{1}{c}{}                                           & \textbf{Healthy}                                         & \textbf{S11}                                    & \multicolumn{1}{c}{71,36\%}                                                         & \multicolumn{1}{c}{\textbf{84,80\%}}                                       & \multicolumn{1}{c}{80,20\%}       & \multicolumn{1}{c}{79,83\%}       & \multicolumn{1}{c}{81,32\%}       & \multicolumn{1}{c}{79,70\%}       & \multicolumn{1}{c}{79,70\%}       & \multicolumn{1}{c}{81,20\%}       & \multicolumn{1}{c}{81,07\%}       & \multicolumn{1}{c}{81,94\%}       & \multicolumn{1}{c}{81,57\%}       & 81,94\%        \\ \cline{2-15} 
\multicolumn{1}{c}{}                                           & \textbf{Healthy}                                         & \textbf{S12}                                    & \multicolumn{1}{c}{\textbf{92,49\%}}                                                         & \multicolumn{1}{c}{87,78\%}                                       & \multicolumn{1}{c}{79,20\%}       & \multicolumn{1}{c}{80,25\%}       & \multicolumn{1}{c}{85,01\%}       & \multicolumn{1}{c}{90,07\%}       & \multicolumn{1}{c}{88,38\%}       & \multicolumn{1}{c}{88,81\%}       & \multicolumn{1}{c}{89,55\%}       & \multicolumn{1}{c}{89,02\%}       & \multicolumn{1}{c}{90,50\%}       & 88,91\%        \\ \cline{2-15} 
\multicolumn{1}{c}{}                                           & \textbf{Healthy}                                         & \textbf{S13}                                    & \multicolumn{1}{c}{\textbf{90,38\%}}                                                         & \multicolumn{1}{c}{88,60\%}                                       & \multicolumn{1}{c}{88,80\%}       & \multicolumn{1}{c}{88,80\%}       & \multicolumn{1}{c}{67,43\%}       & \multicolumn{1}{c}{69,49\%}       & \multicolumn{1}{c}{82,29\%}       & \multicolumn{1}{c}{82,97\%}       & \multicolumn{1}{c}{85,26\%}       & \multicolumn{1}{c}{85,37\%}       & \multicolumn{1}{c}{86,17\%}       & 88,34\%        \\ \cline{2-15} 
\multicolumn{1}{c}{}                                           & \textbf{Healthy}                                         & \textbf{S14}                                    & \multicolumn{1}{c}{\textbf{92,14\%}}                                                         & \multicolumn{1}{c}{89,96\%}                                       & \multicolumn{1}{c}{88,10\%}       & \multicolumn{1}{c}{86,43\%}       & \multicolumn{1}{c}{87,65\%}       & \multicolumn{1}{c}{90,77\%}       & \multicolumn{1}{c}{88,99\%}       & \multicolumn{1}{c}{89,77\%}       & \multicolumn{1}{c}{90,43\%}       & \multicolumn{1}{c}{90,99\%}       & \multicolumn{1}{c}{91,99\%}       & 90,66\%        \\ \cline{2-15} 
\multicolumn{1}{c}{}                                           & \textbf{Healthy}                                         & \textbf{S15}                                    & \multicolumn{1}{c}{\textbf{88,07\%}}                                                         & \multicolumn{1}{c}{85,96\%}                                       & \multicolumn{1}{c}{84,96\%}       & \multicolumn{1}{c}{86,61\%}       & \multicolumn{1}{c}{86,39\%}       & \multicolumn{1}{c}{86,39\%}       & \multicolumn{1}{c}{86,17\%}       & \multicolumn{1}{c}{86,28\%}       & \multicolumn{1}{c}{86,39\%}       & \multicolumn{1}{c}{87,60\%}       & \multicolumn{1}{c}{88,04\%}       & 86,39\%        \\ \cline{2-15} 
\multicolumn{1}{c}{}                                           & \textbf{Hemorrhagic stroke}                                 & \textbf{S16}                                    & \multicolumn{1}{c}{78,71\%}                                                         & \multicolumn{1}{c}{\textbf{80,16\%}}                                       & \multicolumn{1}{c}{73,30\%}       & \multicolumn{1}{c}{78,50\%}       & \multicolumn{1}{c}{66,76\%}       & \multicolumn{1}{c}{67,29\%}       & \multicolumn{1}{c}{69,16\%}       & \multicolumn{1}{c}{70,09\%}       & \multicolumn{1}{c}{71,70\%}       & \multicolumn{1}{c}{75,83\%}       & \multicolumn{1}{c}{74,77\%}       & 75,17\%        \\ \cline{2-15} 
\multicolumn{1}{c}{}                                           & \textbf{Ischemic stroke}                                   & \textbf{S17}                                    & \multicolumn{1}{c}{\textbf{94,38\%}}                                                         & \multicolumn{1}{c}{74,11\%}                                       & \multicolumn{1}{c}{83,44\%}       & \multicolumn{1}{c}{84,20\%}       & \multicolumn{1}{c}{83,82\%}       & \multicolumn{1}{c}{89,63\%}       & \multicolumn{1}{c}{90,52\%}       & \multicolumn{1}{c}{89,51\%}       & \multicolumn{1}{c}{90,90\%}       & \multicolumn{1}{c}{91,53\%}       & \multicolumn{1}{c}{91,53\%}       & 92,29\%        \\ \cline{2-15} 
\multicolumn{1}{c}{}                                           & \textbf{Ischemic stroke}     & \textbf{S18}                                    & \multicolumn{1}{c}{\textbf{91,27\%}}                                                         & \multicolumn{1}{c}{85,70\%}                                       & \multicolumn{1}{c}{84,71\%}       & \multicolumn{1}{c}{85,13\%}       & \multicolumn{1}{c}{87,59\%}       & \multicolumn{1}{c}{86,63\%}       & \multicolumn{1}{c}{88,24\%}       & \multicolumn{1}{c}{89,30\%}       & \multicolumn{1}{c}{89,84\%}       & \multicolumn{1}{c}{89,95\%}       & \multicolumn{1}{c}{89,09\%}       & 90,16\%        

        \\ \hline \multicolumn{1}{l}{} &

      & \textbf{Mean}                                           
      & \multicolumn{1}{c}{\textbf{85,88\%}}                                         & \multicolumn{1}{c}{80,89\%}                                       & \multicolumn{1}{c}{80,71\%}       & \multicolumn{1}{c}{80,43\%}       & \multicolumn{1}{c}{78,86\%}       & \multicolumn{1}{c}{78,13\%}       & \multicolumn{1}{c}{81,97\%}       & \multicolumn{1}{c}{79,55\%}       & \multicolumn{1}{c}{80,18\%}       & \multicolumn{1}{c}{81,88\%}       & \multicolumn{1}{c}{83,71\%}       & 82,36\%        \\\cline{3-15} 
\multicolumn{1}{l}{}                                             &                                           & \textbf{Std.dev}                                        & \multicolumn{1}{c}{\textbf{6,60\%}}                                                          & \multicolumn{1}{c}{7,30\%}                                        & \multicolumn{1}{c}{8,80\%}        & \multicolumn{1}{c}{9,87\%}        & \multicolumn{1}{c}{14,56\%}       & \multicolumn{1}{c}{17,44\%}       & \multicolumn{1}{c}{8,42\%}        & \multicolumn{1}{c}{17,10\%}       & \multicolumn{1}{c}{17,38\%}       & \multicolumn{1}{c}{13,23\%}       & \multicolumn{1}{c}{8,08\%}        & 13,44\%
        
\end{tabular}
\label{tab:Multicenter_raw_data}
\end{table*}

% Table: 1200 sampled Hoffman + UTEC
\begin{table*}[]
\tiny
\caption{Classification accuracy of OE+ME w/ AS (1200 samples)}
\begin{tabular}{cllcccccccccccc}
\hline
\multicolumn{1}{c}{\multirow{2}{*}{\textbf{Dataset}}} & \multicolumn{1}{c}{\multirow{2}{*}{\textbf{Diagnosis}}} & \multicolumn{1}{c}{\multirow{2}{*}{\textbf{Subjects}}} & \multirow{2}{*}{\textbf{Subject-dependent}} & \multirow{2}{*}{\textbf{Subject-independent}} & \multicolumn{10}{c}{\textbf{Subject-adaptive}}                                                                                                                                                                                                                                                                                                             \\ \cline{6-15} 
\multicolumn{1}{c}{}                                           & \multicolumn{1}{c}{}                                    & \multicolumn{1}{c}{}                                   &                                            &                                               & \multicolumn{1}{c}{\textbf{10\%}} & \multicolumn{1}{c}{\textbf{20\%}} & \multicolumn{1}{c}{\textbf{30\%}} & \multicolumn{1}{c}{\textbf{40\%}} & \multicolumn{1}{c}{\textbf{50\%}} & \multicolumn{1}{c}{\textbf{60\%}} & \multicolumn{1}{c}{\textbf{70\%}} & \multicolumn{1}{c}{\textbf{80\%}} & \multicolumn{1}{c}{\textbf{90\%}} & \textbf{100\%} \\ \hline

\multicolumn{1}{c}{\multirow{8}{*}{\textbf{OE}}}          & \textbf{CP}                                              & \textbf{S01}                                   & 82,73\%                                    & 73,07\%                                       & \multicolumn{1}{c}{64,48\%}       & \multicolumn{1}{c}{81,02\%}       & \multicolumn{1}{c}{76,89\%}       & \multicolumn{1}{c}{80,66\%}       & \multicolumn{1}{c}{82,97\%}       & \multicolumn{1}{c}{82,00\%}       & \multicolumn{1}{c}{82,73\%}       & \multicolumn{1}{c}{82,97\%}       & \multicolumn{1}{c}{82,36\%}       & 82,85\%        \\ \cline{2-15} 
\multicolumn{1}{c}{}                                           & \textbf{MS}                                              & \textbf{S02}                                   & 77,16\%                                    & 82,98\%                                       & \multicolumn{1}{c}{69,70\%}       & \multicolumn{1}{c}{82,05\%}       & \multicolumn{1}{c}{82,40\%}       & \multicolumn{1}{c}{82,87\%}       & \multicolumn{1}{c}{82,87\%}       & \multicolumn{1}{c}{82,87\%}       & \multicolumn{1}{c}{82,87\%}       & \multicolumn{1}{c}{82,87\%}       & \multicolumn{1}{c}{82,75\%}       & \textbf{83,33\%}        \\ \cline{2-15} 
\multicolumn{1}{c}{}                                           & \textbf{Late stage ALS}                             & \textbf{S03}                                   & \textbf{90,69\%}                                    & 81,28\%                                       & \multicolumn{1}{c}{82,84\%}       & \multicolumn{1}{c}{83,09\%}       & \multicolumn{1}{c}{82,84\%}       & \multicolumn{1}{c}{83,09\%}       & \multicolumn{1}{c}{83,33\%}       & \multicolumn{1}{c}{82,84\%}       & \multicolumn{1}{c}{82,97\%}       & \multicolumn{1}{c}{83,58\%}       & \multicolumn{1}{c}{82,60\%}       & 85,91\%        \\ \cline{2-15} 
\multicolumn{1}{c}{}                                           & \textbf{TBI \& SCI level C4}                              & \textbf{S04}                                   & 83,21\%                                    & 80,11\%                                       & \multicolumn{1}{c}{80,94\%}       & \multicolumn{1}{c}{81,77\%}       & \multicolumn{1}{c}{81,30\%}       & \multicolumn{1}{c}{83,21\%}       & \multicolumn{1}{c}{82,73\%}       & \multicolumn{1}{c}{82,73\%}       & \multicolumn{1}{c}{82,49\%}       & \multicolumn{1}{c}{82,97\%}       & \multicolumn{1}{c}{83,21\%}       & \textbf{83,69\%}        \\ \cline{2-15} 
\multicolumn{1}{c}{}                                           & \textbf{Healthy}                                         & \textbf{S06}                                   & \textbf{83,82\%}                                    & 79,76\%                                       & \multicolumn{1}{c}{83,33\%}       & \multicolumn{1}{c}{82,61\%}       & \multicolumn{1}{c}{83,09\%}       & \multicolumn{1}{c}{82,85\%}       & \multicolumn{1}{c}{82,97\%}       & \multicolumn{1}{c}{82,73\%}       & \multicolumn{1}{c}{82,97\%}       & \multicolumn{1}{c}{82,61\%}       & \multicolumn{1}{c}{82,61\%}       & 83,09\%        \\ \cline{2-15} 
\multicolumn{1}{c}{}                                           & \textbf{Healthy}                                         & \textbf{S07}                                   & 83,33\%                                    & 54,21\%                                       & \multicolumn{1}{c}{82,40\%}       & \multicolumn{1}{c}{82,17\%}       & \multicolumn{1}{c}{82,52\%}       & \multicolumn{1}{c}{82,98\%}       & \multicolumn{1}{c}{83,10\%}       & \multicolumn{1}{c}{82,40\%}       & \multicolumn{1}{c}{82,52\%}       & \multicolumn{1}{c}{82,52\%}       & \multicolumn{1}{c}{84,27\%}       & 84,27\%        \\ \cline{2-15} 
\multicolumn{1}{c}{}                                           & \textbf{Healthy}                                         & \textbf{S08}                                   & 83,33\%                                    & 73,75\%                                       & \multicolumn{1}{c}{74,69\%}       & \multicolumn{1}{c}{82,10\%}       & \multicolumn{1}{c}{82,47\%}       & \multicolumn{1}{c}{82,84\%}       & \multicolumn{1}{c}{82,22\%}       & \multicolumn{1}{c}{82,35\%}       & \multicolumn{1}{c}{82,35\%}       & \multicolumn{1}{c}{83,21\%}       & \multicolumn{1}{c}{83,09\%}       & \textbf{85,56\%}        \\ \cline{2-15} 
\multicolumn{1}{c}{}                                           & \textbf{Healthy}                                         & \textbf{S09}                                   & \textbf{82,25\%}                                    & 79,92\%                                       & \multicolumn{1}{c}{81,76\%}       & \multicolumn{1}{c}{81,88\%}       & \multicolumn{1}{c}{79,59\%}       & \multicolumn{1}{c}{81,88\%}       & \multicolumn{1}{c}{81,76\%}       & \multicolumn{1}{c}{80,31\%}       & \multicolumn{1}{c}{81,28\%}       & \multicolumn{1}{c}{81,04\%}       & \multicolumn{1}{c}{79,59\%}       & 81,40\%    

\\ \hline

\multicolumn{1}{c}{\multirow{9}{*}{\textit{\textbf{ME}}}}    
& \textbf{Healthy}  & \textbf{S10}          & 
\textbf{97,37\%}                                    & 88,47\%                         & \multicolumn{1}{c}{92,73\%}       & \multicolumn{1}{c}{91,73\%}       & \multicolumn{1}{c}{90,48\%}       & \multicolumn{1}{c}{92,86\%}       & \multicolumn{1}{c}{91,60\%}       & \multicolumn{1}{c}{92,98\%}       & \multicolumn{1}{c}{92,23\%}       & \multicolumn{1}{c}{92,48\%}       & \multicolumn{1}{c}{93,61\%}       & 92,98\%        \\ \cline{2-15} 
\multicolumn{1}{c}{}                                           & \textbf{Healthy}                                         & \textbf{S11}                                    & 79,46\%                                    & \textbf{84,57\%}                                       & \multicolumn{1}{c}{78,81\%}       & \multicolumn{1}{c}{78,55\%}       & \multicolumn{1}{c}{80,88\%}       & \multicolumn{1}{c}{78,04\%}       & \multicolumn{1}{c}{80,75\%}       & \multicolumn{1}{c}{79,33\%}       & \multicolumn{1}{c}{80,23\%}       & \multicolumn{1}{c}{79,46\%}       & \multicolumn{1}{c}{78,81\%}       & 79,97\%        \\ \cline{2-15} 
\multicolumn{1}{c}{}                                           & \textbf{Healthy}                                         & \textbf{S12}                                    & \textbf{93,94\%}                                    & 88,77\%                                       & \multicolumn{1}{c}{81,93\%}       & \multicolumn{1}{c}{83,10\%}       & \multicolumn{1}{c}{85,20\%}       & \multicolumn{1}{c}{86,25\%}       & \multicolumn{1}{c}{87,65\%}       & \multicolumn{1}{c}{88,23\%}       & \multicolumn{1}{c}{89,39\%}       & \multicolumn{1}{c}{89,16\%}       & \multicolumn{1}{c}{91,03\%}       & 88,11\%        \\ \cline{2-15} 
\multicolumn{1}{c}{}                                           & \textbf{Healthy}                                         & \textbf{S13}                                    & 88,97\%                                    & 88,64\%                                       & \multicolumn{1}{c}{\textbf{89,22\%}}       & \multicolumn{1}{c}{62,41\%}       & \multicolumn{1}{c}{74,06\%}       & \multicolumn{1}{c}{80,08\%}       & \multicolumn{1}{c}{83,71\%}       & \multicolumn{1}{c}{83,58\%}       & \multicolumn{1}{c}{83,71\%}       & \multicolumn{1}{c}{86,47\%}       & \multicolumn{1}{c}{88,22\%}       & 88,72\%        \\ \cline{2-15} 
\multicolumn{1}{c}{}                                           & \textbf{Healthy}                                         & \textbf{S14}                                    & \textbf{93,94\%}                                    & 85,90\%                                       & \multicolumn{1}{c}{85,73\%}       & \multicolumn{1}{c}{85,73\%}       & \multicolumn{1}{c}{84,47\%}       & \multicolumn{1}{c}{85,73\%}       & \multicolumn{1}{c}{84,60\%}       & \multicolumn{1}{c}{87,88\%}       & \multicolumn{1}{c}{86,62\%}       & \multicolumn{1}{c}{85,86\%}       & \multicolumn{1}{c}{85,35\%}       & 84,72\%        \\ \cline{2-15} 
\multicolumn{1}{c}{}                                           & \textbf{Healthy}                                         & \textbf{S15}                                    & \textbf{90,80\%}                                    & 86,73\%                                       & \multicolumn{1}{c}{88,43\%}       & \multicolumn{1}{c}{87,81\%}       & \multicolumn{1}{c}{87,44\%}       & \multicolumn{1}{c}{88,06\%}       & \multicolumn{1}{c}{88,43\%}       & \multicolumn{1}{c}{88,43\%}       & \multicolumn{1}{c}{88,68\%}       & \multicolumn{1}{c}{89,05\%}       & \multicolumn{1}{c}{89,18\%}       & 86,57\%        \\ \cline{2-15} 
\multicolumn{1}{c}{}                                           & \textbf{Hemorrhagic stroke}                                 & \textbf{S16}                                    & 78,27\%                                    & \textbf{83,65\%}                                       & \multicolumn{1}{c}{78,93\%}       & \multicolumn{1}{c}{78,67\%}       & \multicolumn{1}{c}{75,07\%}       & \multicolumn{1}{c}{69,60\%}       & \multicolumn{1}{c}{71,60\%}       & \multicolumn{1}{c}{76,13\%}       & \multicolumn{1}{c}{74,00\%}       & \multicolumn{1}{c}{81,33\%}       & \multicolumn{1}{c}{75,33\%}       & 77,20\%        \\ \cline{2-15} 
\multicolumn{1}{c}{}                                           & \textbf{Ischemic stroke}                                   & \textbf{S17}                                    & \textbf{95,67\%}                                    & 82,89\%                                       & \multicolumn{1}{c}{83,46\%}       & \multicolumn{1}{c}{87,79\%}       & \multicolumn{1}{c}{88,68\%}       & \multicolumn{1}{c}{90,59\%}       & \multicolumn{1}{c}{92,11\%}       & \multicolumn{1}{c}{92,75\%}       & \multicolumn{1}{c}{92,62\%}       & \multicolumn{1}{c}{92,24\%}       & \multicolumn{1}{c}{92,37\%}       & 90,97\%        \\ \cline{2-15} 
\multicolumn{1}{c}{}                                           & \textbf{Ischemic stroke}                                   & \textbf{S18}                                    & \textbf{92,72\%}                                    & 83,57\%                                       & \multicolumn{1}{c}{84,62\%}       & \multicolumn{1}{c}{86,15\%}       & \multicolumn{1}{c}{87,21\%}       & \multicolumn{1}{c}{87,68\%}       & \multicolumn{1}{c}{87,56\%}       & \multicolumn{1}{c}{88,97\%}       & \multicolumn{1}{c}{89,20\%}       & \multicolumn{1}{c}{88,50\%}       & \multicolumn{1}{c}{87,79\%}       & 89,32\%

\\ \hline

\multicolumn{1}{l}{}                                             &                                                          & \textbf{Mean}                                           & \textbf{86,92\%}                                    & 81,07\%                                       & \multicolumn{1}{c}{81,41\%}       & \multicolumn{1}{c}{82,27\%}       & \multicolumn{1}{c}{82,62\%}       & \multicolumn{1}{c}{83,49\%}       & \multicolumn{1}{c}{84,12\%}       & \multicolumn{1}{c}{84,50\%}       & \multicolumn{1}{c}{84,52\%}       & \multicolumn{1}{c}{85,08\%}       & \multicolumn{1}{c}{84,83\%}       & 85,22\%        \\ \cline{3-15} 
\multicolumn{1}{l}{}                                             &                                                          & \textbf{Std.dev}                                        & 6,45\%                                    & 8,30\%                                        & \multicolumn{1}{c}{6,90\%}        & \multicolumn{1}{c}{6,14\%}        & \multicolumn{1}{c}{4,54\%}        & \multicolumn{1}{c}{5,22\%}        & \multicolumn{1}{c}{4,66\%}        & \multicolumn{1}{c}{4,63\%}        & \multicolumn{1}{c}{4,72\%}        & \multicolumn{1}{c}{\textbf{3,93\%}}        & \multicolumn{1}{c}{4,95\%}        & 4,01\%         \\ \cline{3-15} 
\end{tabular}
\label{tab:Multicenter_sampling}
\end{table*}

\subsection{Computation time} %\cite{Zhdanov2019} 
\begin{figure}[htpb!]
     \centering
     \includegraphics[width = 0.33\textwidth]{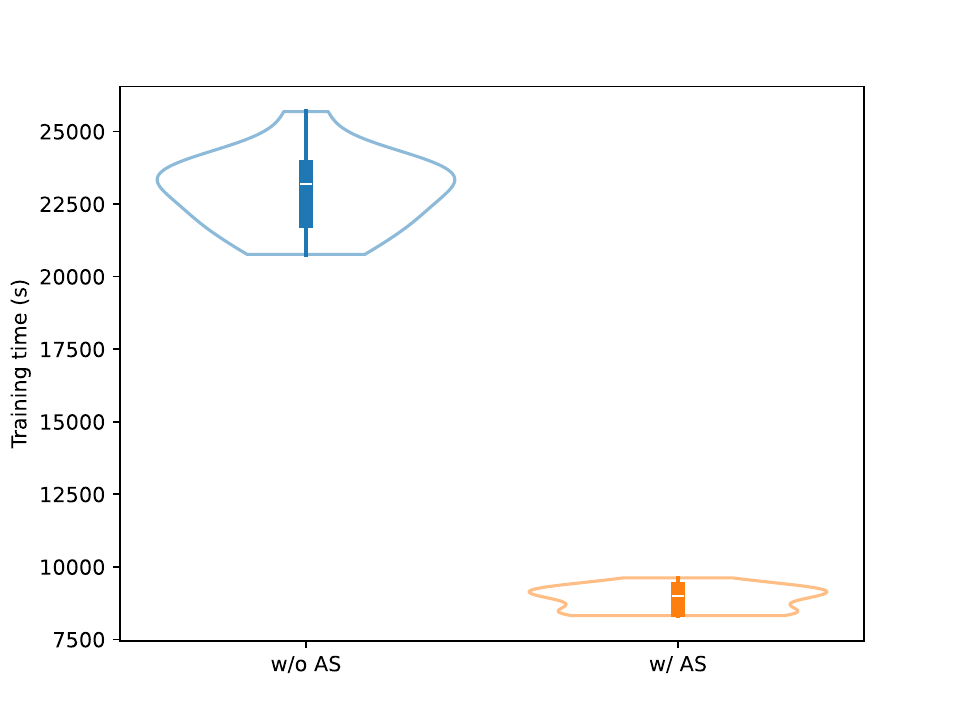}
     \caption{Boxplot of training time for different sets of data with significant statistical differences ($p=1.526e-05$) with Wilcoxon test and p-value=0.001. Raw data approx has 3000 samples.}
     %\caption{Boxplot of training time for different sets of data %with significant statistical different (p=9.907e-30) with %Wilcoxon test and p-value=0.001}. Each set was tested to be a normal distribution with Kolmogorov-Smirnov test with p-values of 0.9945(raw data) and 0.6018(1200 samples). Additionally, a t-test was run to determine their statistical difference, and the outcome rejected the null hypothesis with p-value = 6.8597e-18.
     \label{fig:training_time}
\end{figure}

Reducing samples with Active Sampling (AS) aims to significantly cut the CPU training time (in seconds) for Deep4net. Table \ref{tab:training_time} demonstrates this reduction by comparing average CPU training times with and without AS, using 1200 samples across both datasets (OE and IE).
With AS, the average CPU time was $22896.44 \pm 1292.02$ s (Avg. $\pm$ SD), while without AS, it averaged $8889.75\pm 419.14$ s (Avg. $\pm$ SD), resulting in a substantial 61.17\% reduction. Notably, the average CPU time did not follow a normal distribution. A nonparametric Wilcoxon signed-rank test was conducted to analyze the median differences between w/o AS and w/ AS, yielding highly significant p-values $(p < 0.01)$, specifically $1.526 \times 10^{-5}$.
Subject 14 achieved the minimum CPU training time without AS, while subject 8 had the best reduction, with a remarkable 64.46\% reduction in CPU training time (in seconds) with AS.
 % (8318.52/23405.82=0.35540391236 
 % (23405.82-8318.52)/23405.82*100= 64.459608764
 %process with 0.70.
%On the other hand, the average CPU time (sec) reduction training process was 0.39.
% 8889.75/22896.44=0.38825904813%.825 
\begin{table}[h!]
\scriptsize
\caption{CPU time (sec) comparison between training with w/o AS and w/ AS in sampling 1200}
\scalebox{.78}{
\begin{tabular}{ccccc}
\hline
 &  &  & \multicolumn{2}{c}{\textbf{Training time (s)}} \\ \cline{4-5} 
\multirow{-2}{*}{\textbf{Dataset}} & \multirow{-2}{*}{\textbf{Diagnosis}} & \multirow{-2}{*}{\textbf{Subjects}} & w/o AS & \multicolumn{1}{l}{w/ AS (1200 samples)} \\ \hline
 & \textbf{CP} & \textbf{S01} & 23207,00 & 9112,38 \\ \cline{2-5} 
 & \textbf{MS} & \textbf{S02} & 22418,51 & 8991,86 \\ \cline{2-5} 
 & \textbf{Late stage ALS} & \textbf{S03} & 25683,09 & 9168,53 \\ \cline{2-5} 
 & \textbf{TBI \& SCI level C4} & \textbf{S04} & 23781,20 & 9025,06 \\ \cline{2-5} 
 & \textbf{Healthy} & \textbf{S06} & 23190,75 & 8406,21 \\ \cline{2-5} 
 & \textbf{Healthy} & \textbf{S07} & 24190,93 & 8337,90 \\ \cline{2-5} 
 & \textbf{Healthy} & \textbf{S08} & 23405,82 & \textbf{8318,52} \\ \cline{2-5} 
\multirow{-8}{*}{\textbf{OE}} & \textbf{Healthy} & \textbf{S09} & 22.819,26 & 9274,83
  \\ \hline
  & \textbf{Healthy} & \textbf{S10} & 21554,44 & 9260,78 \\ \cline{2-5} 
 & \textbf{Healthy} & \textbf{S11} & 23621,02 & 9387,30 \\ \cline{2-5} 
 & \textbf{Healthy} & \textbf{S12} & 21907,92 & 9233,98 \\ \cline{2-5} 
 & \textbf{Healthy} & \textbf{S13} & 20784,10 & 8499,96 \\ \cline{2-5} 
 & \textbf{Healthy} & \textbf{S14} & \textbf{20763,63} & 8495,82 \\ \cline{2-5} 
 & \textbf{Healthy} & \textbf{S15} &  23935,36 &  8353,36 \\ \cline{2-5} 
 & \textbf{Hemorrhagic stroke} & \textbf{S16} & 23847,56 & 8851,04 \\ \cline{2-5} 
 & \textbf{Ischemic stroke} & \textbf{S17} & 22315,38 & 9618,95 \\ \cline{2-5} 
\multirow{-9}{*}{\textbf{ME}} & \textbf{Ischemic stroke} & \textbf{S18} & 21813,49 & 8789,19
  \\ \hline
\multicolumn{1}{l}{} & \multicolumn{1}{l}{} & \textbf{Mean} & 22896,44 & 8889,75 \\ \cline{3-5}
\multicolumn{1}{l}{} & \multicolumn{1}{l}{} & \textbf{Std.dev} & 1292,02 & 419,14 \\ \cline{3-5}
\end{tabular}}
\label{tab:training_time}
\end{table}
We used violin plots to visually demonstrate the distribution of samples before and after applying Active Sampling (AS). Figure \ref{fig:training_time} presents a violin plot depicting the median, upper adjacent value, and lower adjacent value for the average CPU time (sec) before and after applying AS for all 17 subjects. The violin plot for the dataset without AS shows higher median and quartile values. However, upon applying Active Sampling (AS), the plot displays a downward shift in the entire distribution, indicating a significant reduction in CPU time (sec) during the training process. It's crucial to emphasize that implementing Active Sampling (AS) on each subject's data leads to a substantial reduction in CPU training time (s).
%0.4

\subsection{Bitrates Analysis}

\begin{figure}
   \centering
     \begin{subfigure}[b]{0.24\textwidth}
         \centering
         \includegraphics[width = \textwidth]{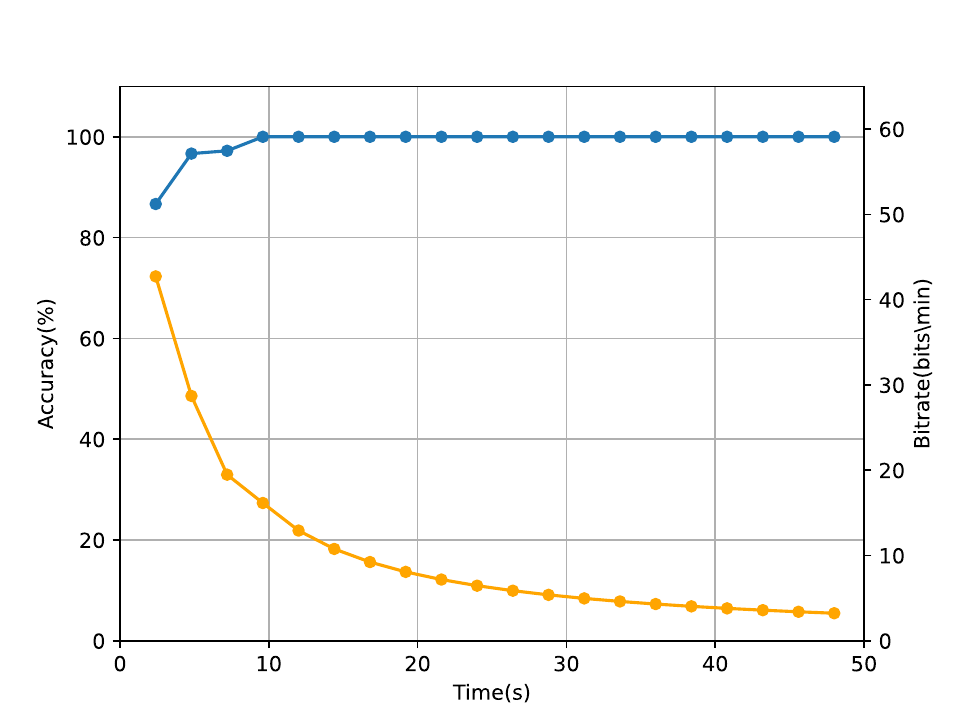}
         \caption{w/o AS}
         \label{fig:bitrate_raw}
     \end{subfigure}
     \hfill
     \begin{subfigure}[b]{0.24\textwidth}
         \centering
         \includegraphics[width = \textwidth]{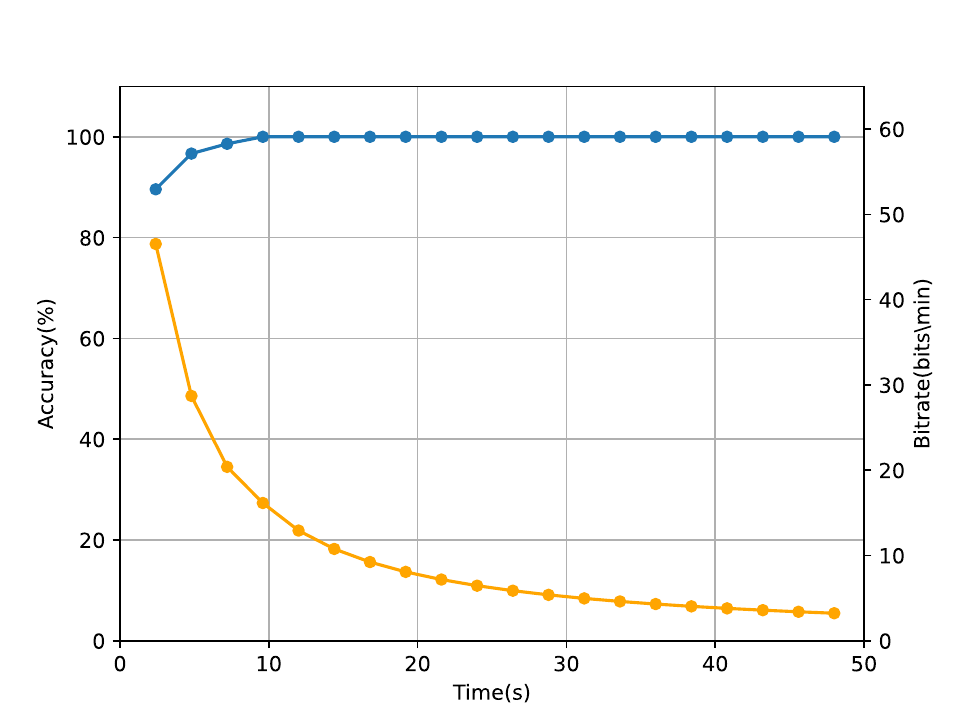}
         \caption{w/ AS (1200 samples)}
         \label{fig:bitrate_sampling}
     \end{subfigure}
     \caption{Bitrate results of S10 from OE+ME}
     \label{fig:bitrate_results}
\end{figure}

In Fig. \ref{fig:bitrate_raw}, averaged classification accuracy and their corresponding bitrates are plotted against the time needed for the number of blocks to make a decision. The bitrates were calculated using the definition
of \cite{WOLPAW_2002} utilizing the average accuracy curves. To demonstrate the feasibility of Active Sampling (AS), we select subject 10, corresponding to the experimental scheme of OE+ME.  If bitrate is used as a performance measure, differences between OE+ME w/ AS and OE+ME w/o AS can be observed.   For instance, OE+ME w/ AS outperformed OE+ME w/o AS for the bitrate from the first block, where w/ AS reported roughly 80 bits/min. As we can see, there is a close link to maximum bitrate and high classification accuracy for small numbers of blocks. 

Differences between OE+ME w/ AS and OE+ME w/o AS can also be observed for averaged classification accuracy. The classification accuracy of OE+ME w/o AS increases more slowly than  OE+ME  w/ AS, as we can see before the 10 seconds. Both metrics demonstrate that Active Sampling (AS) increases the bitrate and averaged classification accuracy using less computational time.

% Best Adpative for Multi center expansion

\section{Discussion} \label{discussion}
%Comparison with other methodology
% Results comparison with previous works

A key challenge in Brain-Computer Interface (BCI) research is achieving high classification accuracy with minimal calibration data while ensuring maximum generalizability. Transfer learning has been a common approach to address this, but fine-tuning often leads to over-fitting of the target class and irrelevant feature emphasis, even if the original model is highly robust. The variability of EEG data across individuals and experiments further complicates this. Our study, unique in its multi-subject, multi-country, and multi-hardware scope, introduces a novel adaptation transfer method. A novel method of adaptation transfer called ATL using Active Sampling (AS), based on Dense PDS is proposed and evaluated in such a scenario. Despite the data diversity, our deep learning model showed enhanced classification accuracy and reduced CPU training time. As detailed in Section V, our approach and active sampling method achieved 85.22\% accuracy in subject-independent, heterogeneous conditions, surpassing other methods in real-world scenarios and demonstrating faster training times. 

The stability of dense PDS was confirmed by conducting multiple repetitions of training and testing the model in Adaptive Transfer Learning (ATL) for subject-independent classification using the experimental scheme from OE+ME. The results are in Table S1 from Section S.II from the supplementary material. The mean classification accuracy results of each repetition show that Dense PDS is stable and consistent, indicating its robustness and reliability. Therefore, to compare the performance of Dense PDS, we included additional Active Sampling (AS) techniques based on Vanilla PDS, Easy PDS, and Anneal PDS \cite{Zhang2018} in Adaptive Transfer Learning (ATL) methods using the raw data of each subject for subject-independent classification using the experimental scheme from OE+ME. The results can be found in Table S2 in Section S.III of the supplementary material. The results show that PDS Dense is the best choice, achieving the highest classification accuracy. Furthermore, to compare the performance of  Adaptive Transfer Learning (ATL) using Deep4Net, over the OE+ME with AS for subject-independent classification, we applied Instance-based Transfer Learning (ITL) using Weighted k-Nearest Neighbors (Weighted K-NN). The results can be found in Table S3 in Section S.V of the supplementary material. The results demonstrate that the mean classification accuracy of Deep4Net outperformed Weighted K-NN, showcasing its superiority. Thus, this work offers new insights into achieving optimal generalization in P300-BCI systems.

\section{Conclusion} \label{conclusion}
This study introduces an adaptive transfer methodology with sampling selection for P300-BCI, showing notable performance in subject-independent and multi-centre contexts. The methodology leverages a diverse dataset from various countries, times, and hardware, enhancing P300 detection robustness and minimizing overfitting. Notably, it significantly reduces training time, enabling quick BCI-P300 system calibration and reactivation within a minute. These findings advance the generalizability of non-invasive P300 BCI systems. Future applications of these techniques to other BCI paradigms are planned, with the possibility of integrating adaptive transfer learning with generative models for improved accuracy. This later methodology, however, requires further meticulous exploration. In the future, we plan to explore integrations of this transfer methodology in other machine learning and deep learning techniques \cite{gupta2022gentle, kiani2022towards, andreu2011real, kiani2019effective, al2020deep, andreu2021single}.

\section*{Acknowledgment}
The authors thank all participants who took part in this investigation and anonymous reviewers for valuable feedback.

% https://ieeexplore.ieee.org/abstract/document/9672118/authors#authors

%The authors thank all the participants who took part in this study. The authors thank the anonymous reviewers for their valuable feedback and suggestions.

%The preferred spelling of the word ``acknowledgment'' in American English is  without an ``e'' after the ``g.'' Use the singular heading even if you have  many acknowledgments. Avoid expressions such as ``One of us (S.B.A.) would like to thank $\ldots$ .'' Instead, write ``F. A. Author thanks $\ldots$ .'' In most  cases, sponsor and financial support acknowledgments are placed in the  unnumbered footnote on the first page, not here.

\bibliographystyle{ieeetr} 
\bibliography{MAIN}

\end{document}